\documentclass[preprint,12pt,authoryear]{elsarticle}

\usepackage{anyfontsize} \usepackage{setspace} \usepackage{lineno} \usepackage{graphicx} \usepackage{multirow} \usepackage{amsmath,amssymb,amsfonts} \usepackage{amsthm} \usepackage{mathrsfs} \usepackage[title]{appendix} \usepackage{xcolor} \usepackage{textcomp} \usepackage{manyfoot} \usepackage{booktabs} \usepackage{array} \usepackage{tabularx} \usepackage{algorithm} \usepackage{algorithmicx} \usepackage{algpseudocode} \usepackage{listings} \usepackage{hyperref} \usepackage[capitalise]{cleveref}
\AtBeginDocument{\hypersetup{hypertexnames=false}}
\theoremstyle{plain}   \theoremstyle{definition}   \theoremstyle{remark} 

\journal{Journal of Hydrology}

\begin{document}

\singlespacing

\begin{frontmatter}

    \title{
        Design-Life Levels for Flood Extremes: A Dependence-Aware Block-Maxima Workflow for Severity and Persistence
    }

    \author[inst1]{Tuoyuan Cheng\corref{cor1}}
    \ead{tuoyuan.cheng@nus.edu.sg}
    \cortext[cor1]{Corresponding author}

    \author[inst2]{Xiao Peng}
    \ead{xp53@nus.edu.sg}

    \author[inst3]{Achmad Choiruddin}
    \ead{choiruddin@its.ac.id}

    \author[inst2]{Xiaogang He}
    \ead{hexg@nus.edu.sg}

    \author[inst1,inst4]{Kan Chen}
    \ead{kan.chen@nus.edu.sg}

    \affiliation[inst1]{
        organization={Risk Management Institute, National University of Singapore},
        city={Singapore},
        postcode={119244},
        country={Singapore}
    }
    \affiliation[inst2]{
        organization={Department of Civil and Environmental Engineering, National University of Singapore},
        city={Singapore},
        postcode={117576},
        country={Singapore}
    }
    \affiliation[inst3]{
        organization={Department of Statistics, Institut Teknologi Sepuluh Nopember},
        city={Surabaya},
        postcode={60111},
        country={Indonesia}
    }
    \affiliation[inst4]{
        organization={Department of Mathematics, National University of Singapore},
        city={Singapore},
        postcode={119076},
        country={Singapore}
    }

    \begin{abstract}
        Flood-risk assessment concerns the magnitude of maximum discharge over a design life and the clustering of extreme flows in time.
        Limited record length, extremal clustering, and pre-asymptotic behavior complicate inference from daily streamflow and flood-impact records, yet severity estimation, persistence assessment, and design-life levels are often handled separately.
        We develop a dependence-aware block-maxima workflow for approximately stationary records with regularly varying upper tails.
        The severity branch estimates the extreme value index (EVI) from sliding block-maximum quantile scaling using data-adaptive plateau selection and covariance-aware feasible generalized least squares (FGLS).
        The persistence branch pools native block-maxima extremal-index paths over a stable block-size window to characterize extremal clustering.
        Design-life levels are then derived on the relevant observation clock, with the extremal index retained as a complementary persistence descriptor.
        In synthetic short-record benchmarks, median-sliding-FGLS achieves the lowest grid-average Winkler score for nominal 95\% EVI confidence intervals among the methods compared.
            For the extremal index, FGLS pooling of sliding-block Berghaus--B{\"u}cher estimates lowers mean Winkler scores relative to pooled ordinary least squares in all 84 scenarios and to the native estimator in 81.
        Applications to Texas and Florida streamflow and National Flood Insurance Program building-claim records show stronger extremal clustering in streamflow and steeper block-maximum scaling in insured losses on the active-day clock.
    \end{abstract}


\end{frontmatter}

\section{Introduction}\label{sec:introduction}

Flood-frequency analysis often reports annual exceedance probabilities and return levels, whereas design decisions concern finite service lives.
For flood design and risk management, a complementary question is therefore how large the maximum discharge, flood impact, or insured loss may become over a specified horizon, such as 10, 25, or 50 years \citep{rootzen_katz_design_life_2013,salas_obeysekera_2014,read_vogel_2015,choiruddin2024algorithms}.
Dependent records also require an assessment of how strongly extreme observations cluster in time \citep{ferro_segers_2003,moloney_overview_2019}.
These questions connect extreme-value estimation to infrastructure reliability, operational stress, and flood-loss management \citep{smolka_natural_2006,smith_us_2013}, while avoiding the interpretation of return-period labels as literal waiting times \citep{serinaldi_dismissing_2015}.

Classical extreme value theory (EVT) provides the limiting framework for marginal tail behavior and for extremes of dependent sequences \citep{leadbetter_extremes_1983,katz_statistics_2002,davison_statistics_2015,gomes_extreme_2015}.
In hydrologic applications, however, inference is shaped by short records, serial dependence with extremal clustering, and slow convergence toward limiting approximations \citep{merz_thieken_uncertainty_2005,metzger_flood_2020,clarkson_importance_2023,belzile_modelers_2023}.
We focus on a univariate workflow for flood-related streamflow and impact records.
Tail severity is summarized through the extreme value index (EVI) \(\xi\) \citep{hill_simple_1975,smith1987estimating,schultze1996least,meerschaert1998simple,fedotenkov_review_2020,jochem_oorschot_tail_2023}; clustering is summarized through the extremal index (EI) \(\theta\) \citep{moloney_overview_2019,ferreira_clustering_2024,ferreira_extremal_2023}; and fitted extreme-value models provide return levels and design-life levels for hydrologic design.
The practical challenge is to estimate these quantities from short records with uncertainty assessments that account for dependence.

Block-maxima (BM) methods are a natural starting point for design-life analysis because the target is a maximum over a finite horizon \citep{ana_ferreira_block_2015,davison_statistics_2015,coles_introduction_2001}.
The difficulty is how to use the record efficiently without understating uncertainty.
Sliding blocks provide more overlapping maxima than disjoint blocks and can improve statistical efficiency, provided that dependence between overlapping maxima and across block sizes is accounted for \citep{bucher_segers_sliding_2018,northrop_efficient_2015}.
Inference must therefore account for covariance among block-maximum quantile estimates across block sizes.

Return-period terminology adds a communication problem because it is often read as a waiting-time statement rather than as a finite-horizon risk statement \citep{serinaldi_dismissing_2015}.
Design-life levels instead describe the distribution of the maximum over a specified horizon, which aligns directly with hydrologic design and flood-risk assessment \citep{rootzen_katz_design_life_2013,fawcett_sea_surge_2016,fawcett_bayesian_2018,chen_measuring_2022}.
Recent hydrologic design work likewise frames uncertainty and reliability over planning horizons rather than only through fixed annual exceedance probabilities \citep{xiong_robustness_2024,slater_nonstationary_2021,shabestanipour_risk-based_2024,barbaux_integrating_2025}.

We develop a dependence-aware block-maxima workflow for short, serially dependent hydrologic records \citep{ana_ferreira_block_2015,northrop_efficient_2015,wager_subsampling_2014}.
The severity branch estimates the EVI from sliding block-maximum quantile scaling, using adaptive plateau selection and feasible generalized least squares (FGLS) with bootstrap covariance estimates.
The persistence branch pools transformed block-maxima EI estimates by FGLS over a selected stable window \citep{northrop_efficient_2015,berghaus_weak_2018,moloney_overview_2019,ferreira_extremal_2023}.
Design-life levels are derived from the severity fit on the chosen observation clock \citep{rootzen_katz_design_life_2013}.
Separate short-record benchmarks assess point accuracy and interval performance for EVI and EI estimation.

The workflow is applied to four longer-record flood-related series: Texas and Florida streamflow and Texas and Florida National Flood Insurance Program (NFIP) building-claim records \citep{czajkowski_economic_2016,nelson_mercer_pluvial_2025}.
Together, these applications show how severity escalation and persistence can be interpreted across adjacent parts of the flood-risk chain without reducing them to a single aggregate risk score.

The remainder of the paper is organized as follows.
Section~\ref{sec:methodology} presents the proposed methodology.
Section~\ref{sec:simulation} reports the synthetic benchmark evidence.
Section~\ref{sec:applications} applies the workflow to streamflow and NFIP claims.
Section~\ref{sec:discussion} discusses hydrologic implications, limitations, and conclusions.

\section{Methodology}\label{sec:methodology}
This section presents a dependence-aware workflow for short, serially dependent hydrologic records with heavy upper tails \citep{leadbetter_extremes_1983,davison_statistics_2015}.
The goal is finite-sample inference for design-life levels and extremal persistence, not a general comparison of all EVT domains of attraction.
We work with a Fr\'echet-domain approximation and assume approximate stationarity over the retained analysis period after preprocessing, as is common in hydrologic extremes applications \citep{clarkson_importance_2023,belzile_modelers_2023,zwiers_climate_2013}.
The application screening assesses compatibility with the assumed Fr\'echet domain.

Starting from a univariate source record, the workflow returns three interpretable quantities: an estimate of the extreme value index (EVI) \(\widehat\xi\), an estimate of the extremal index (EI) \(\widehat\theta\), and the derived design-life level \(D_\tau(T)\), where \(T\) is the design life in years and \(\tau\in(0,1)\) is the non-exceedance probability for the maximum over that period \citep{rootzen_katz_design_life_2013,serinaldi_dismissing_2015}.
\Cref{fig:workflow} summarizes that logic.
The severity branch maps the input series to a block-maximum quantile scaling relation and fits a covariance-aware log--log model to estimate the EVI \(\widehat\xi\), from which the design-life levels are derived.
The persistence branch constructs native BM EI paths from the Northrop and Berghaus--B{\"u}cher (BB) estimators \citep{northrop_efficient_2015,berghaus_weak_2018}, maps them to the common transformed scale \(z_b=\log(1/\widehat{\theta}_b)\), and pools the transformed path over an EI stable window to obtain the EI estimate \(\widehat\theta\).
The following subsections distinguish what is assumed, what is estimated directly, what is pooled, and what is subsequently derived for interpretation.

\begin{figure}[htbp]
    \centering
    \setlength{\fboxsep}{10pt}
    \fbox{
        \begin{minipage}{0.88\textwidth}
            \centering
            \begin{tabular}{p{0.40\textwidth}p{0.40\textwidth}}
                \multicolumn{1}{c}{\textbf{Severity branch}}                                                                & \multicolumn{1}{c}{\textbf{Persistence branch}}                                                                                            \\
                series on the severity clock                                             & series on the persistence clock                                                                         \\
                \(\downarrow\) sliding block maxima across candidate block sizes                                            & \(\downarrow\) native sliding BM EI paths (BB/Northrop) across candidate block sizes                                                       \\
                \(\downarrow\) log-quantile path (median)                  & \(\downarrow\) transformed EI path \(z_b=\log(1/\widehat\theta_b)\)                                                                        \\
                \(\downarrow\) score-based plateau selection                                                                & \(\downarrow\) score-based EI stable window selection                                    \\
                \(\downarrow\) super-block bootstrap covariance & \(\downarrow\) circular block-bootstrap covariance of the transformed EI path \\
                \(\downarrow\) FGLS fit for EVI \(\widehat\xi\)                     & \(\downarrow\) FGLS pooling for EI \(\widehat\theta\)                                          \\
                \(\downarrow\) derived design-life levels \(D_\tau(T)\)                                                     & \(\downarrow\) estimated limiting mean cluster size \(1/\widehat\theta\)              \\
            \end{tabular}
        \end{minipage}
    }
    \caption{
        Dependence-aware block-maxima workflow for severity and persistence.
        The severity branch estimates the EVI \(\widehat\xi\) and derives design-life levels \(D_\tau(T)\) for non-exceedance probability \(\tau\) over design life \(T\).
        The persistence branch estimates the EI \(\widehat\theta\); \(1/\widehat\theta\) estimates the limiting mean number of upper-tail exceedances per cluster.
        The branches use a common source record, with input series and candidate block-size grids defined separately on their respective observation clocks.
    }
    \label{fig:workflow}
\end{figure}

\subsection{Setup, scope, and inferential targets}
For each inferential branch, let \(X_1,\ldots,X_N\) denote the univariate hydrologic or flood-impact series obtained after its specified preprocessing.
Here \(N\) is the record length, \(t\) indexes observations on the specified clock, and \(\log\) denotes the natural logarithm.
The branches share a source record, but their preprocessing, observation clocks, and candidate block-size grids need not coincide.
Because design-life questions concern maxima over finite horizons, we analyze block maxima over a grid of block sizes rather than fixing one block size in advance \citep{rootzen_katz_design_life_2013,salas_obeysekera_2014,read_vogel_2015}.
For block size \(b\), define the sliding block maxima
\begin{equation}\label{eq:blockmax}
    M_{b,t}=\max\{X_t,\ldots,X_{t+b-1}\}, \qquad t=1,\ldots,N-b+1,
\end{equation}
and let \(M_b\) denote a generic block maximum of size \(b\), with distribution \(F_b\).
Sliding blocks advance by one observation, yielding \(n_b^{\mathrm{sl}}=N-b+1\) maxima.
    Disjoint blocks start at \(t=1,1+b,\ldots\), yielding \(n_b^{\mathrm{dis}}=\lfloor N/b\rfloor\) maxima.
    Sliding blocks are the default; disjoint maxima need not be independent.
For a quantile level \(\tau \in (0,1)\), write
\begin{equation}\label{eq:blockquantile}
    Q_{\tau}(M_b)=F_b^{\leftarrow}(\tau)=\inf\{x:F_b(x)\ge \tau\},
\end{equation}
where \(F_b^{\leftarrow}\) is the generalized inverse of \(F_b\).

For either scheme, the empirical block-maximum quantile is
\begin{equation}\label{eq:empirical_block_quantile}
    \widehat Q_{\tau}(M_b)=\inf\left\{x:\widehat F_b(x)\ge \tau\right\},
\end{equation}
where \(\widehat F_b\) is the empirical distribution of the extracted block maxima.
A hat denotes a sample-based estimate.

The severity branch fits the block-maximum quantile path across \(b\).
    Its primary target is \(Q_{0.5}(M_b)\), the median block maximum, which is used because it is empirically stable in the short-record benchmark in Section~\ref{sec:simulation}.
    Mean and mode summaries are retained only as robustness baselines.
    The mode baseline is computed with a smoothed kernel-density surrogate; details are given in \ref{app:severity-bootstrap-backbone}.

Serial dependence, overlapping windows, and reuse of observations across block sizes are accounted for in the covariance estimation below.

For calendar-time analysis of zero-flow or rainfall records, genuine zero observations are retained when forming blocks; missing observations must not be coded as zeros.
    Zero-valued block maxima are included when estimating the quantile at each block size.
    Only block sizes with strictly positive quantile estimates are eligible for the log--log severity fit.
    Bootstrap covariance estimation retains only replicates with positive quantile estimates at every fitted block size.
    Frequent all-zero blocks can therefore leave too few usable block sizes or bootstrap replicates for reliable inference.

Both branches are evaluated on a geometrically spaced grid \(b_1,\ldots,b_K\) of \(K\) candidate block sizes, but the resulting block-size paths have different roles.
For severity, the task is to identify a plateau over which the log-quantile path is approximately linear in log block size, so that \(\xi\) can be estimated.
For persistence, the task is to identify an EI stable window over which the transformed BM EI path, \(z_b=\log(1/\widehat\theta_b)\), can be pooled to estimate \(\theta\).
The workflow uses explicit window-selection criteria to make these choices reproducible.

\subsection{The severity branch: block-quantile scaling}
The severity branch asks how the block-maximum quantile \(Q_{\tau}(M_b)\) changes as the block size \(b\) increases.
In the Fr\'echet domain of attraction, regular variation motivates an approximate log--log scaling relation for block-maximum quantiles over intermediate block sizes \citep{leadbetter_extremes_1983,davison_statistics_2015,clauset_power-law_2009}.
We use the working relation
\begin{equation}\label{eq:evi_scaling}
    \log Q_{\tau}(M_b) \approx \alpha_{\tau} + \xi \log b,
\end{equation}
where \(\alpha_\tau\) is the quantile-specific intercept and the extreme value index \(\xi\) is the slope.
Larger values of \(\xi\) imply faster growth of block maxima as the design life expands.
The primary EVI estimate is the slope \(\widehat{\xi}\) from the sliding-block fit to the median block maximum \(Q_{0.5}(M_b)\).

    The median is taken over block maxima, rather than over individual observations in the input series.
    Here \(b\) is the number of observations per block on the chosen observation clock, and \(M_b\) is the corresponding block maximum.
    Let \(F\) denote the marginal distribution of an individual observation \(X_t\) in the stationary series.
    The normalizing scale \(a_b=F^{\leftarrow}(1-1/b)\) is the marginal \(1-1/b\) quantile and has the same units as \(X_t\).
    Under regular variation with \(\xi>0\) and the extremal limit conditions for a positive extremal index \(\theta\),
    \(\Pr(M_b/a_b\le x)\to\exp\{-\theta x^{-1/\xi}\}\) for any fixed dimensionless threshold \(x>0\) as \(b\to\infty\) \citep{leadbetter_extremes_1983,northrop_efficient_2015}.
    Consequently, for any fixed \(\tau\in(0,1)\),
    \[
        Q_\tau(M_b)\sim a_b\left(\frac{\theta}{-\log\tau}\right)^\xi,
        \qquad a_b=b^\xi\ell(b),
    \]
    where \(\ell\) is a positive slowly varying function, satisfying \(\ell(cb)/\ell(b)\to1\) as \(b\to\infty\) for every fixed \(c>0\).
    Thus, fixed interior quantiles, including the median, share the scaling exponent \(\xi\); \(\tau\) and \(\theta\) enter the multiplicative factor.
    Equation~\eqref{eq:evi_scaling} is a local approximation over windows where \(\log\ell(b)\) varies little, consistent with the block-maxima scaling argument of \citet{stoev_michailidis_2010}.
    The median requires no finite moments.

The log--log relation is not expected to hold across all block sizes in a short hydrologic record.
Small blocks may retain pre-asymptotic curvature, whereas blocks that are large relative to the record length provide few effectively independent maxima, increasing finite-sample uncertainty and potentially weakening asymptotic approximations \citep{stoev_michailidis_2010,bucher_segers_sliding_2018}.
The workflow therefore constructs the log-quantile path over the full grid, then fits the scaling relation only over a data-adaptive plateau.
This window is used to balance curvature and variance in estimating the EVI.
The construction is consistent with EVT work on intermediate sequences, penultimate approximations, and regression-style EVI estimation \citep{smith1987estimating,schultze1996least,gomes_extreme_2015}.

\subsubsection{Plateau selection}
Write \(x_j=\log b_j\) and \(y_j=\log\widehat Q_{0.5}(M_{b_j})\) for the log-quantile path, which traces the log median block maximum against log block size.
Let \([j_1,j_2)\) denote a candidate window on the log-quantile path.
    The window contains indices \(j=j_1,\ldots,j_2-1\), so \(j_2-j_1\) is its number of grid points.
    Following common EVT usage, a plateau is a block-size range over which local log--log scaling is sufficiently stable for EVI estimation.
    Here stability means local linearity of the log-quantile path together with limited variation in adjacent slopes.
    This choice follows established stability-based approaches to scale and threshold selection in EVT.
        In the max-spectrum method, \citet{stoev_michailidis_2010} select a block-scale range by monitoring changes in the fitted slope as the range expands.
        The Reiss--Thomas threshold-selection rule instead minimizes a penalized mean deviation along a tail-index path \citep{neves_fraga_alves_2004}.
        Equations~\eqref{eq:plateau_score} and~\eqref{eq:curvature_score} define a criterion constructed here to implement these principles for the block-quantile path.
    Candidate windows are scored by
    \begin{equation}\label{eq:plateau_score}
        \mathcal S(j_1,j_2)
        =
        \frac{\mathrm{MSE}(j_1,j_2) + \lambda_{\mathrm{curv}}\kappa(j_1,j_2)}{\sqrt{j_2-j_1}},
    \end{equation}
    where \(\mathrm{MSE}(j_1,j_2)\) denotes the mean squared error, computed here as the mean squared residual from the local linear fit, and
    \begin{equation}\label{eq:curvature_score}
        \kappa(j_1,j_2)
        =
        \frac{1}{j_2-j_1-2}\sum_{j=j_1+1}^{j_2-2}\left|\Delta \widehat m_j\right|
    \end{equation}
    penalizes changes in adjacent local slopes.
    Here \(\widehat m_j=(y_{j+1}-y_j)/(x_{j+1}-x_j)\), \(\Delta \widehat m_j=\widehat m_j-\widehat m_{j-1}\), and \(\lambda_{\mathrm{curv}}\) is the curvature-penalty weight, fixed across all analyses.
    The residual term measures overall departure from linearity, while the curvature term detects changes in local slope that a single fit can obscure.
        The square-root length factor reduces the preference for short, incidentally straight windows; it is not an effective-sample-size correction.
        The weight \(\lambda_{\mathrm{curv}}=2\) and this length factor are fixed empirical settings; \Cref{tab:implementation-defaults} lists the implementation defaults.
        The nearby-window ranges in \Cref{tab:application-selection-sensitivity} describe sensitivity to window choice under this fixed criterion.

    \subsubsection{Dependence-aware covariance estimation and FGLS inference}\label{sec:fgls-inference}
    After a plateau has been selected, inference must account for dependence along the fitted block-size path.
    The regression inputs \(\log \widehat Q_{\tau}(M_b)\) are correlated because sliding blocks overlap and neighboring block sizes reuse much of the same record.
    A conventional least-squares fit can estimate the working log--log relation, but its uncertainty estimates do not reflect this covariance across block sizes.

    We estimate the covariance matrix of the retained log block-maximum quantile path with a super-block bootstrap for block maxima, and then use that estimate in feasible generalized least squares (FGLS).
    Each super-block is a contiguous segment longer than the block sizes used for the fit.
        For sliding blocks, we compute circular maxima within each segment and use the same super-block draws across block sizes to reconstruct the log-quantile path.
        Resampling these maxima collections together retains their internal dependence and provides joint bootstrap replicates for covariance estimation.
        This adapts the circular-block-maxima construction of \citet{bucher_staud_bootstrap_2026}, who establish consistency under suitable conditions and show that naive blockwise resampling of sliding block maxima can be inconsistent.
        Here, its use across multiple block sizes is a finite-sample covariance approximation under stationarity and short-range dependence; its adequacy depends on the length and number of super-blocks.
        Details are given in \ref{app:severity-bootstrap-backbone}.

    Let \(Y_k=\log \widehat Q_{\tau}(M_{b_k})\) and \(x_k=\log b_k\).
    Over a selected plateau with \(m\) retained block sizes, relabelled \(k=1,\ldots,m\), write
    \begin{equation}\label{eq:fgls_model}
        Y = X\beta + \varepsilon, \qquad
        X=
        \begin{bmatrix}
            1      & x_1    \\
            \vdots & \vdots \\
            1      & x_m
        \end{bmatrix},
        \qquad
        \beta=
        \begin{bmatrix}
            \alpha_{\tau} \\
            \xi
        \end{bmatrix}.
    \end{equation}
    The vector \(Y=(Y_1,\ldots,Y_m)^\top\) collects the retained log-quantile estimates, \(X\) is the \(m\times2\) design matrix, and \(\beta\) contains the intercept and slope; \({}^\top\) denotes transposition.
    Here \(\varepsilon\) denotes a centered approximation error for the retained log-quantile path, with covariance matrix \(\Sigma\) across block sizes.
    Each super-block bootstrap replicate yields a vector \(Y^{*(r)}\) of log-quantile estimates across the retained block sizes.
    The superscript \(*(r)\) labels bootstrap replicate \(r=1,\ldots,R\), where \(R\) is the number of bootstrap replicates. In both FGLS branches, \(R\) is selected separately for each fit using the finite-bootstrap Monte Carlo standard error (MCSE) rule in \ref{app:implementation-defaults}.
    We estimate this covariance by
    \begin{equation}\label{eq:bootstrap_cov}
        \widehat\Sigma
        =
        \frac{1}{R-1}
        \sum_{r=1}^{R}
        \left(Y^{*(r)}-\bar Y^*\right)\left(Y^{*(r)}-\bar Y^*\right)^{\!\top},
    \end{equation}
    where \(\bar Y^*=R^{-1}\sum_{r=1}^R Y^{*(r)}\) is the mean bootstrap path.
    This is the usual corrected sums-of-products covariance estimator applied to the bootstrap path draws \citep{welford_note_1962}.
    Before inversion, we apply diagonal shrinkage and a small ridge term:
        \begin{equation}\label{eq:covariance_regularization}
            \begin{aligned}
                \widehat\Sigma_{\mathrm{reg}}
                 & = (1-\delta)\widehat\Sigma
                +\delta\,\operatorname{diag}(\widehat\Sigma)
                +r_{\mathrm{ridge}}I_p,                                                               \\
                r_{\mathrm{ridge}}
                 & =\max\left\{10^{-8}\frac{\operatorname{tr}(\widehat\Sigma)}{p},\,10^{-12}\right\}.
            \end{aligned}
        \end{equation}
        Here \(p=m\) is the covariance dimension, \(I_p\) is the identity matrix, \(\operatorname{diag}(\widehat\Sigma)\) retains only the diagonal of \(\widehat\Sigma\), and \(\operatorname{tr}\) denotes the trace.
        The shrinkage factor \(\delta\in[0,1]\) attenuates off-diagonal covariances while preserving the diagonal before the ridge addition.
        Both FGLS branches use \(\delta=0.37\) as a common empirical shrinkage setting in the reported benchmarks and applications.
        This fixed computational choice is not an optimum selected from the sensitivity analysis.
        The severity branch uses the FGLS estimator
    \begin{equation}\label{eq:fgls_est}
        \widehat\beta_{\mathrm{FGLS}}
        =
        (X^{\top}\widehat\Sigma_{\mathrm{reg}}^{-1}X)^{-1}X^{\top}\widehat\Sigma_{\mathrm{reg}}^{-1}Y.
    \end{equation}

    The working covariance of the fitted coefficients is \(\widehat V=(X^\top\widehat\Sigma_{\mathrm{reg}}^{-1}X)^{-1}\), with \(\widehat V_{22}\) denoting the estimated slope variance.
        The approximate 95\% Wald confidence interval (CI) for the EVI \(\xi\) is \(\widehat\xi\pm1.96\sqrt{\widehat V_{22}}\), computed on the selected plateau using the fixed regularization settings.
    Interval performance is evaluated in the short-record benchmark in Section~\ref{sec:simulation}.
    The same regularized bootstrap covariance matrix is therefore used for both covariance-aware reweighting and interval construction.
    \ref{app:severity-bootstrap-backbone} describes the bootstrap construction, \Cref{tab:implementation-defaults} lists the fixed regularization defaults, and \Cref{fig:benchmark-evi-shrinkage} in \ref{app:shrinkage-sensitivity} reports the shrinkage-grid sensitivity check.
    The remaining implementation notes are collected in \ref{app:implementation-defaults} and \ref{app:algorithms}.

    \subsubsection{Estimation of design-life levels}
    The primary severity output is the design-life level.
    In flood-frequency practice, fitted levels are often reported through annual exceedance probabilities or equivalent return-period labels.
    Here we report the finite-horizon quantity directly: the \(\tau\)-quantile of the maximum over a \(T\)-year horizon on the chosen observation clock.
    Formally,
    \begin{equation}\label{eq:design_life_level}
        D_{\tau}(T)=Q_{\tau}(M_{b_T}), \qquad b_T=\lceil\lambda T\rceil,
    \end{equation}
    where \(\lambda\) is the observation rate, in observations per year, on the application-specific clock.
    The ceiling converts the stated horizon to an integer block size.
    For the applications below, \(\lambda\) corresponds to the calendar-day rate for streamflow and to the observed active-day rate for NFIP severity; the two clocks are discussed further in Section~\ref{sec:applications}.

    Under the scaling relation in \eqref{eq:evi_scaling}, the estimator is
    \begin{equation}\label{eq:design_life_level_est}
        \widehat D_{\tau}(T)=\exp\{\widehat\alpha_{\tau}+\widehat\xi\log b_T\}.
    \end{equation}
    Here \(\widehat\alpha_\tau\) is the estimated intercept for quantile level \(\tau\) in the log--log scaling relation.
    When \(b_T\) exceeds the largest block size in the selected plateau, this estimate extrapolates beyond the fitted block-size range.
    The primary design-life curve uses \(\tau=0.50\), giving the median design-life level.

    For the reported median design-life level, let \(\widehat\beta=(\widehat\alpha_{0.5},\widehat\xi)^\top\), let \(\widehat V\) be the fitted coefficient covariance matrix defined in Section~\ref{sec:fgls-inference}, and set \(a_T=(1,\log b_T)^\top\).
        The estimated log-scale variance is
        \[
            \widehat s_T^2=a_T^\top\widehat V a_T
            =\widehat V_{11}+2\log b_T\,\widehat V_{12}
            +(\log b_T)^2\widehat V_{22}.
        \]
        Here, \(\widehat V_{11}\) and \(\widehat V_{22}\) are the estimated intercept and slope variances, and \(\widehat V_{12}\) is their covariance.
        With standard error \(\widehat s_T\), the pointwise 95\% confidence interval is
        \begin{equation}\label{eq:design_life_ci}
            \left[
                \exp\{a_T^\top\widehat\beta-1.96\,\widehat s_T\},\,
                \exp\{a_T^\top\widehat\beta+1.96\,\widehat s_T\}
                \right].
        \end{equation}
        This normal-approximation interval quantifies uncertainty in the fitted median with the observation rate, selected window, and regularization settings held fixed.
        It is not a prediction interval for the future maximum and excludes uncertainty in window selection, future dynamics, and the validity of the scaling extrapolation.

    Motivated by the common asymptotic scaling exponent in \eqref{eq:evi_scaling}, we construct design-life levels at \(\tau'=0.90,0.95,0.99\) using the median-fit slope \(\widehat\xi\) and the set of block sizes \(\mathcal P\) in its selected plateau.

    Only the quantile-specific intercept is re-estimated:
    \begin{equation}\label{eq:shared_xi_intercept}
        \widehat\alpha_{\tau'}
        =
        \frac{1}{|\mathcal P|}
        \sum_{b\in\mathcal P}
        \left\{\log \widehat Q_{\tau'}(M_b)-\widehat\xi\log b\right\}.
    \end{equation}
    The cardinality \(|\mathcal P|\) is the number of retained block sizes.
    This shared-slope fit is a bias--variance compromise for short records, where empirical block-maxima quantiles at \(\tau=0.95\) or \(0.99\) can be too noisy to support separate plateau selection and slope estimation.

    \subsubsection{Interpretation of design-life levels}\label{sec:design-life-interpretation}
    All reported design-life levels are conditional on the selected plateau and on the stationary working model over the analysis period.
    These design-life levels complement annual exceedance probabilities and return levels.
    If \(M_T\) denotes the maximum over a \(T\)-year horizon and \(z\) is a candidate design threshold, the design-life statement is
    \begin{equation}\label{eq:design_life_probability}
        \Pr(M_T \le z)=\tau,
    \end{equation}
    where \(\tau\) is the selected non-exceedance probability for the horizon maximum.
    Under the idealized approximation of independent yearly maxima with annual exceedance probability \(p\), the corresponding probability of at least one exceedance over \(T\) years is
    \begin{equation}\label{eq:encounter_probability}
        \Pr(M_T > z)=1-(1-p)^T.
    \end{equation}
    This interpretation follows the design-life formulation of \citet{rootzen_katz_design_life_2013} and related risk- and reliability-based approaches to hydrologic design \citep{salas_obeysekera_2014,read_vogel_2015}.

    \textit{Practical interpretation.}
        Under the fitted scaling relation, an EVI of \(0.65\) implies that extending a horizon from 25 to 50 years increases the corresponding maximum-discharge quantile by a factor of \(2^{0.65}\approx1.57\), for the same non-exceedance probability and observation clock, provided that extrapolation is defensible.
        The EVI alone does not determine an absolute design discharge; the fitted intercept, target probability, and uncertainty are also needed.
        In particular, \(D_{0.5}(50)\) has a model-implied 50\% probability of being exceeded over 50 years; it is neither a 50-year return level nor, by itself, an engineering design criterion.

    \subsection{The persistence branch: pooled EI paths}
    The persistence branch estimates the extremal index \(\theta\) to quantify clustering of extremes.
    Under Leadbetter-type mixing and anti-clustering conditions, \(\theta\in(0,1]\) describes clustering of extremes, and \(1/\theta\) can be interpreted as a limiting mean cluster size \citep{leadbetter_extremes_1983,moloney_overview_2019}.
Cluster size counts upper-tail exceedances, meaning observations above a high threshold, in the limit as that threshold approaches the upper endpoint \citep{ferro_segers_2003}.
    The BM estimators infer this quantity from block maxima without explicitly identifying exceedance clusters.
For each block size, BM EI estimates are computed using the Northrop and Berghaus--B{\"u}cher estimators \citep{northrop_efficient_2015,berghaus_weak_2018}.
The resulting EI path is mapped to
\begin{equation}\label{eq:ei_z_scale}
    z_b=\log(1/\widehat{\theta}_b),
\end{equation}
and the transformed path is pooled over an automatically selected EI stable window.
Here \(\widehat\theta_b\) is the native EI estimate at block size \(b\).
The reciprocal scale retains the limiting mean cluster-size interpretation, while the logarithm gives a nonnegative additive scale with the monotone back-transformation \(\widehat\theta=\exp(-z)\).

\subsubsection{EI stable window pooling on the transformed EI path}
The EI stable window search on the \(z_b\) path parallels plateau selection, but targets a different structure.
The severity branch estimates a local slope, whereas the persistence branch seeks a block-size window over which the transformed EI path is approximately constant.
Let \(\mathcal B_{\mathrm{stab}}\) denote the set of block sizes in the selected EI stable window.
For \(b\in\mathcal B_{\mathrm{stab}}\), we use the intercept-only model
\begin{equation}\label{eq:pooled_z_model}
    z_b = \mu + \eta_b,
\end{equation}
where \(\mu\) is the stable level on the log reciprocal-extremal-index scale and \(\eta_b\) is the deviation from that level at block size \(b\).
This specification assumes stability only within the selected window, not across the full block-size range.

We estimate the covariance of the retained \(z_b\) path by a circular block bootstrap of the input series \citep{politis_romano_circular_1992}.
    Each resample concatenates fixed-length blocks of consecutive observations, with end-to-start wrapping.
    This retains within-block dependence and avoids undersampling observations near the record boundaries.
    Recomputing the empirical distribution and all retained EI estimates on each of the \(R\) resamples provides a joint covariance estimate across block sizes.
    The adaptive stopping rule for \(R\) is given in \ref{app:implementation-defaults}.
    We use this as a finite-sample covariance approximation for stationary series with short-range dependence; its adequacy depends on the resampling block length.
The unconstrained pooled FGLS estimate is
\begin{equation}\label{eq:pooled_z_gls}
    \widetilde\mu
    =
    \left(\mathbf 1^{\top}\widehat\Sigma_{z,\mathrm{reg}}^{-1}\mathbf 1\right)^{-1}
    \mathbf 1^{\top}\widehat\Sigma_{z,\mathrm{reg}}^{-1} z,
\end{equation}
where \(\widehat\Sigma_{z,\mathrm{reg}}\) is obtained from \(\widehat\Sigma_z\) using \Cref{eq:covariance_regularization}, with the same fixed \(\delta=0.37\) and \(p=|\mathcal B_{\mathrm{stab}}|\).
This regularization acts on the covariance used for pooling across block sizes; it does not alter the native estimates \(\widehat\theta_b\).
Here \(z=(z_b)_{b\in\mathcal B_{\mathrm{stab}}}\) collects the retained transformed estimates, \(\mathbf1\) is a vector of ones of the same length, and \(\widehat\Sigma_z\) estimates their covariance.
We impose the parameter bound by setting \(\widehat\mu=\max\{0,\widetilde\mu\}\).
The extremal-index estimate is then
\begin{equation}\label{eq:theta_backtransform}
    \widehat\theta = \exp(-\widehat\mu),
\end{equation}
with \(1/\widehat\theta\) reported as the estimated limiting mean cluster size.
Let \(\widehat s_\mu\) denote the fitted standard error on the \(z\)-scale; for FGLS, \(\widehat s_\mu^2=(\mathbf1^\top\widehat\Sigma_{z,\mathrm{reg}}^{-1}\mathbf1)^{-1}\).
    The approximate 95\% confidence interval is obtained by back-transforming the bounded \(z\)-scale endpoints:
    \begin{equation}\label{eq:theta_ci}
        \left[
            \exp\{-\widehat\mu-1.96\,\widehat s_\mu\},\,
            \exp\{-\max(0,\widehat\mu-1.96\,\widehat s_\mu)\}
            \right].
    \end{equation}

Implementation defaults for EI stable window selection and covariance regularization are listed in \Cref{tab:implementation-defaults}; \Cref{fig:benchmark-ei-shrinkage} in \ref{app:shrinkage-sensitivity} reports the shrinkage-grid sensitivity check.
Threshold-based EI estimators, including K-gaps and Ferro--Segers, are retained as external baselines \citep{ferreira_clustering_2024,ferreira_extremal_2023}.
In the applications, \(\widehat\theta\) is interpreted as a 
descriptor of short-range extremal clustering that complements the severity-side design-life levels.

\textit{Practical interpretation.}
    An EI of \(0.05\) indicates strong extremal clustering and, under the limiting interpretation, a mean cluster size of about \(1/0.05=20\) upper-tail exceedances.
    For daily records, these observations need not be consecutive, so this is not an estimate of a 20-day flood duration.
    A small EI indicates that extremes tend to cluster; event-duration and inter-event-gap diagnostics are needed before translating this summary into reservoir operations or emergency-response schedules.

\section{Synthetic benchmark}\label{sec:simulation}
The synthetic benchmark evaluates the workflow, not only the point accuracy of individual EVI or EI estimators.
This distinction is important because both literatures contain many competing estimators, and existing reviews show that no method dominates uniformly across dependence structures and tail regimes \citep{fedotenkov_review_2020,ferreira_extremal_2023}.
We use two short-record benchmarks with a common design: a fixed length of \(N=365\), interpreted as a stylized daily record, and three dependent heavy-tailed process families for each branch.
Each scenario uses 100 independent simulation replicates.
This setting retains the original observation order and provides a demanding test of block-size selection and interval performance under dependence.

The applications in Section~\ref{sec:applications} use longer flood-related records.
The benchmark nevertheless targets the short-record setting because tail inference remains information-limited even in longer series: large block sizes, high block-quantile targets, serial dependence, and active-day filtering for claim severity all reduce the effective number of extremes.
It is therefore used as a stress test for the proposed inferential workflow before the longer-record applications are considered.

\subsection{Benchmark design}

\subsubsection{Data-generating processes and parameter grids}
We use separate short-record benchmarks because the severity and persistence branches target different parameters.
The EVI benchmark varies \(\xi \in \{0.01,0.03,0.10,0.30,1.0,3.0,10.0\}\) across representative dependence levels \(\theta \in \{0.01,0.10,0.50,1.0\}\).
The EI benchmark places greater resolution on dependence, using
\(\theta \in \{0.10,0.15,0.25,0.40,0.60,0.80,1.0\}\) together with
\(\xi \in \{0.01,0.50,1.0,5.0\}\).
In each benchmark, the data are generated from three dependent heavy-tailed families: a Fr\'echet max-autoregressive (max-AR) process, a moving-maxima process with \(q=99\), and a Pareto additive autoregressive process of order one (AR(1)).

Operationally, these families are
\begin{equation}\label{eq:benchmark_dgps}
    X_t=\max(\phi X_{t-1},Z_t), \qquad
    X_t=\max_{0\le j\le q}\phi^j Z_{t-j}\ \ (q=99), \qquad
    X_t=\phi X_{t-1}+Z_t,
\end{equation}
with \(0\le \phi<1\) and independent positive innovations \(Z_t\).
The coefficient \(\phi\) controls propagation of innovations, and \(q\) is the largest lag in the moving-maxima model.
The max-AR and moving-maxima families use Fr\'echet innovations, whereas the additive AR(1) family uses Pareto innovations.
They represent max-propagation, finite-range moving clustering, and linear accumulation of heavy-tailed shocks, respectively \citep{zhang_peng_idowu_2016,basrak_segers_2009}.
The most extreme tail settings are included as numerical stress cases rather than as realistic hydrologic models.
Thus, the two branches share a parallel benchmark design, but not a single common \((\xi,\theta)\) parameter grid.

\subsubsection{Benchmark estimators and comparison design}
The comparison structure has two layers.
First, within the BM framework, we compare ordinary least squares (OLS) and FGLS variants applied to the same severity and persistence targets.
OLS provides a reference fit to the selected block-size path, whereas FGLS evaluates the effect of using the bootstrap-estimated covariance across block sizes.
These within-BM comparisons isolate the role of block extraction, path construction, and covariance-aware weighting without changing the inferential target.

For severity, the within-BM benchmark varies the block scheme (sliding or disjoint), the block-summary target (median, mean, or mode), and the regression weighting (OLS or FGLS) within a common block-maxima structure.
These BM regressions report Wald intervals for the fitted slope \(\xi\).
The primary median-sliding-FGLS workflow is then compared with external baselines from other methodological classes: Hill, Pickands, and the Dekkers--Einmahl--de Haan (DEdH) moment estimator as raw-sample threshold-based EVI estimators, and max-spectrum as a block-maxima-style scaling baseline.
In the main benchmark, these external methods retain their native asymptotic Gaussian or Wald intervals \citep{hill_simple_1975,pickands_1975,dekkers_einmahl_dehaan_1989,stoev_michailidis_2010,fedotenkov_review_2020}.

For persistence, the within-BM benchmark varies the path construction, block scheme, and OLS/FGLS pooling within the pooled-BM structure.
The pooled BM EI workflows report log-scale Wald intervals after pooling the transformed path \(z_b=\log(1/\widehat\theta_b)\).
They are compared with threshold-based EI estimators (K-gaps and Ferro--Segers) and native fixed-\(b\) BM estimators (Northrop and Berghaus--B{\"u}cher).
The external methods again retain their native interval constructions: bounded Wald intervals for Ferro--Segers \citep{ferro_segers_2003}, profile-likelihood intervals for K-gaps \citep{suveges_davison_2010}, adjusted profile-likelihood intervals for Northrop's estimator \citep{northrop_efficient_2015,chandler_bate_2007}, and bounded Wald intervals for the BB estimator \citep{berghaus_weak_2018}.

This distinction matters for interpretation.
Within-BM comparisons support the strongest attribution claims because the target is held fixed.
Comparisons with external baselines instead assess practical competitiveness across methods with different tuning parameters, target constructions, and interval conventions.
External baselines use pre-specified, method-specific tuning rules that are applied consistently across scenarios.
\ref{app:benchmark-baselines} summarizes the baseline definitions, tuning rules, and interval conventions; \ref{app:benchmark-core-tables} and \ref{app:shrinkage-sensitivity} provide the remaining benchmark conventions and sensitivity checks.

\subsubsection{Evaluation metrics}
The benchmark reports two complementary metrics: the Winkler interval score for interval performance and median absolute percentage error (APE) for point accuracy.
Let \(\psi\in\{\xi,\theta\}\) denote the parameter being evaluated and \(\psi_0\) its true benchmark value.
Interval performance is summarized by the Winkler score for nominal 95\% confidence intervals, which combines interval width with a penalty for noncoverage \citep{winkler_interval_1972,gneiting_raftery_scoring_2007}:
\begin{equation}\label{eq:winkler_metric}
    W_{0.05}(L,U;\psi_0)
        =
        (U-L)
        +
        40(L-\psi_0)_+
        +
        40(\psi_0-U)_+.
\end{equation}
Here, \(L\) and \(U\) are the endpoints of a nominal 95\% confidence interval; lower scores indicate better interval performance.
The notation \((a)_+=\max(a,0)\) denotes the positive part.
We use the mean Winkler score as the primary interval criterion, averaging over valid simulation replicates within each scenario. The grid-average Winkler score is the equally weighted mean of these scenario-level means over the stated parameter grid. Scores are computed on the original parameter scale without normalization.
    Winkler score bars indicate \(\pm\) one standard error (SE) of the mean across simulation replicates, computed as the sample standard deviation of the scores divided by the square root of the number of valid scored fits. They are distinct from parameter confidence intervals and from the finite-bootstrap MCSE used to select the number of bootstrap replicates \(R\) (\ref{app:implementation-defaults}).

Point accuracy alone does not assess interval performance: narrow intervals can under-cover, while broad intervals may lack precision. We therefore interpret Winkler scores alongside empirical coverage.

For an estimate \(\widehat\psi\) of \(\psi\), point accuracy is summarized by
\begin{equation}\label{eq:ape_metric}
    \mathrm{APE}=\frac{|\widehat\psi-\psi_0|}{|\psi_0|}.
\end{equation}
APE measures point accuracy on a relative scale and is reported as a fraction: an APE of 0.55 corresponds to 55\%.
The median limits the influence of unusually large relative errors. APE bars span the 25th--75th percentiles across valid simulation replicates, with the median as the central estimate.
At \(\xi=0.01\), the near-zero denominator can magnify relative errors, so APE is treated as a secondary diagnostic.

\subsection{EVI benchmark}
\subsubsection{Within-BM severity comparison}
\begin{figure}[p]
    \makebox[\textwidth][c]{%
        \begin{minipage}{1.25\textwidth}
            \centering
            \includegraphics[width=\linewidth,height=0.82\textheight,keepaspectratio]{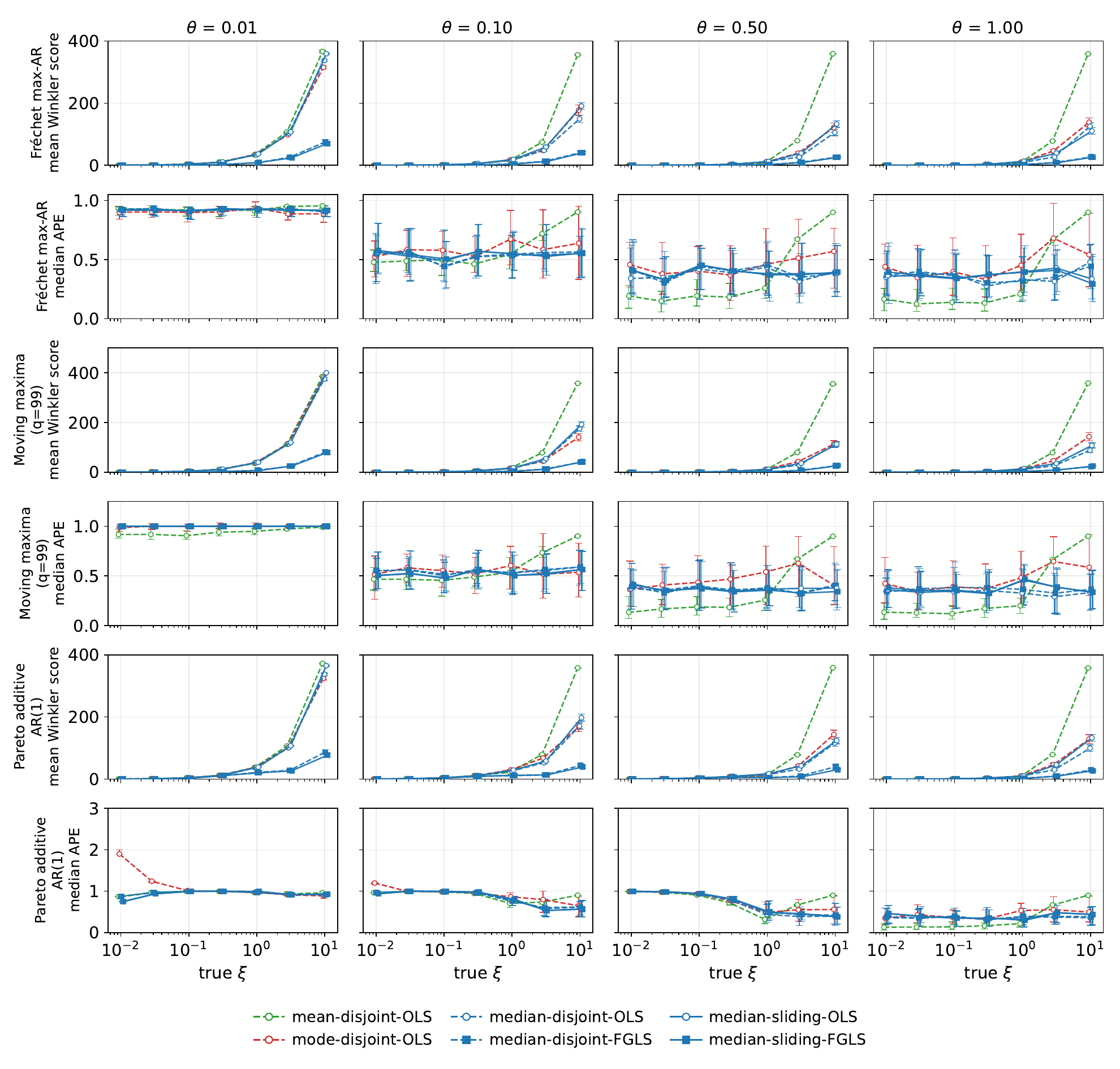}
            \caption{
                Within-block-maxima EVI benchmark.
                Panels summarize mean Winkler score for 95\% confidence intervals (\(\pm1\) SE of the mean) and median absolute percentage error (APE, expressed as a fraction; interquartile range) over the EVI benchmark with \(\xi \in \{0.01,0.03,0.10,0.30,1.0,3.0,10.0\}\), \(\theta \in \{0.01,0.10,0.50,1.0\}\), and three dependent heavy-tailed families: Fr\'echet max-AR, moving-maxima \(q=99\), and Pareto additive AR(1).
                Lower values indicate better interval performance or point accuracy.
                The comparison varies block scheme, block-summary target, and regression weighting within the severity branch.
            }
            \label{fig:benchmark-evi-main}
        \end{minipage}}
\end{figure}

\Cref{fig:benchmark-evi-main} summarizes the within-BM EVI benchmark across the block-maxima severity regressions.
Among these within-BM variants, the median-sliding-FGLS configuration has the lowest grid-average Winkler score.
At \(\theta=0.10\), median-sliding-FGLS lowers the Winkler score averaged over the \(\xi\) grid in each process family, with little change in median APE (\Cref{tab:benchmark-evi-summary}). For example, the Fr\'echet max-AR score falls from \(39.55\) under OLS to \(8.32\) under FGLS, while median APE remains \(0.55\).

Across the severity benchmark grid, all within-BM scenario winners use FGLS, with median-sliding-FGLS attaining the lowest mean score in \(27\) of \(84\) scenarios.

Relative to \(\delta=0\), diagonal shrinkage substantially reduces the grid-average Winkler score, while median APE changes little across positive shrinkage values. The score-minimizing value on the examined grid differs among process families (\Cref{fig:benchmark-evi-shrinkage}).

\subsubsection{External baseline comparison}
\begin{figure}[p]
    \makebox[\textwidth][c]{%
        \begin{minipage}{1.25\textwidth}
            \centering
            \includegraphics[width=\linewidth,height=0.82\textheight,keepaspectratio]{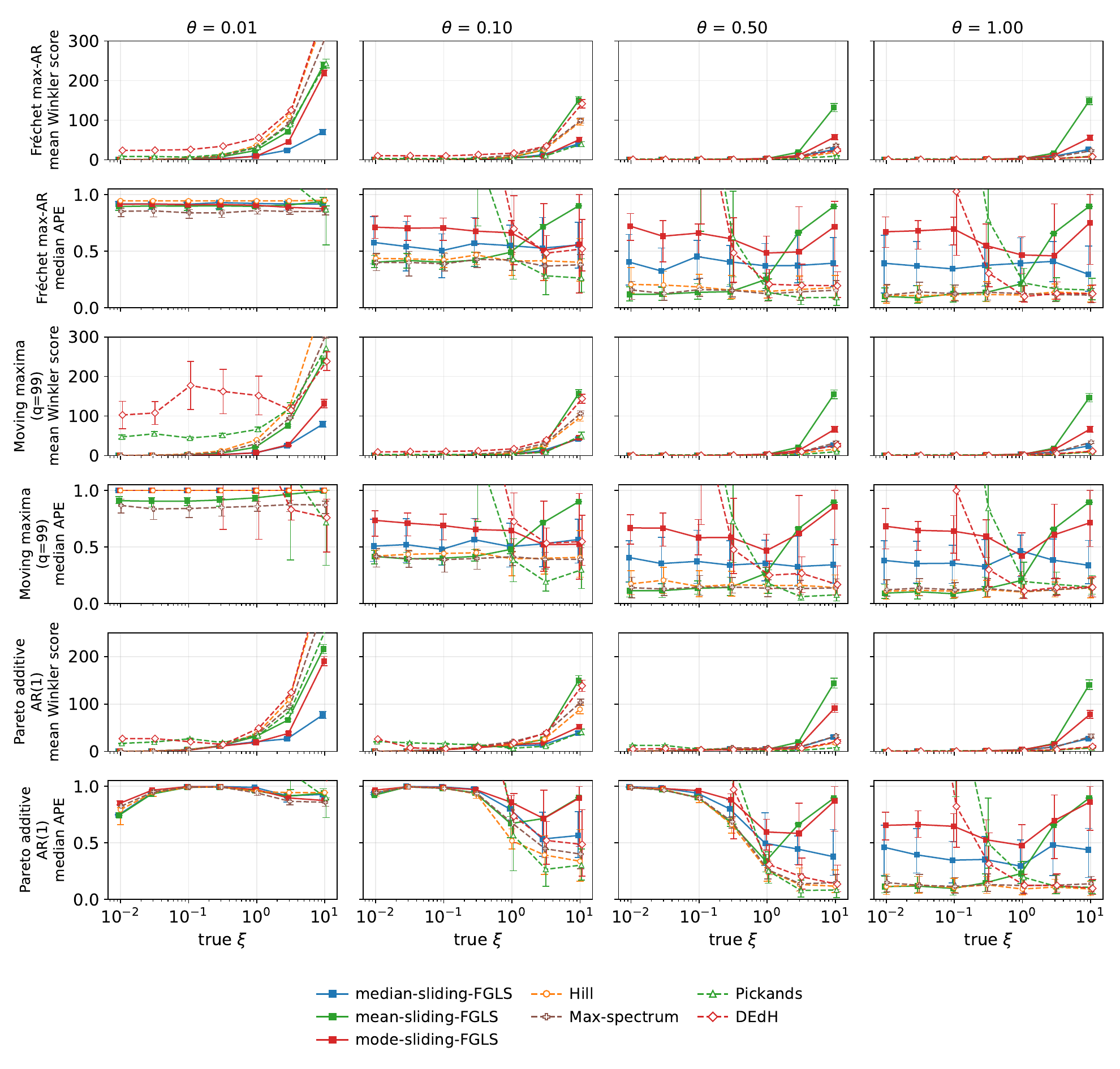}
            \caption{
                EVI comparison with external baseline estimators.
                Panels compare sliding-block FGLS variants based on median, mean, and mode block-maximum summaries with Hill, max-spectrum, Pickands, and DEdH estimators over the same EVI benchmark.
                External baselines are shown with their native asymptotic interval constructions.
                Winkler scores use 95\% confidence intervals; bars show \(\pm1\) SE of the mean. APE is a fraction, shown as the median and interquartile range.
                Vertical limits are set from the internal BM methods; external values outside those limits are clipped for display only and remain in the numerical summaries.
                }
            \label{fig:benchmark-evi-targets}
        \end{minipage}}
\end{figure}

Among the seven methods in \Cref{fig:benchmark-evi-targets}, median-sliding-FGLS never ranks last by mean Winkler score in any of the \(84\) scenarios; each of the other six methods ranks last at least once.

Among the sliding-block FGLS variants, mean Winkler scores rise more steeply for the mean and mode targets at large \(\xi\).

Hill and max-spectrum often have lower APE than median-sliding-FGLS, especially under weaker dependence.
Among the seven methods in \Cref{fig:benchmark-evi-targets}, Hill attains the lowest mean Winkler score in \(22\) of the \(84\) severity scenarios, followed by median-sliding-FGLS in \(19\).
    Median-sliding-FGLS nevertheless has the lowest grid-average Winkler score, \(9.49\), compared with \(25.81\) for Hill.
    These two summaries answer different questions: the number of scenario wins does not measure the magnitude of the losses in the remaining scenarios, and an unnormalized grid-average Winkler score gives greater influence to scenarios with larger interval widths and noncoverage penalties.

At \(\theta=0.10\), median-sliding-FGLS has lower Winkler scores averaged over the \(\xi\) grid than Hill and max-spectrum in each process family (\Cref{tab:benchmark-evi-summary}). Hill and max-spectrum have smaller median APE for max-AR and moving-maxima processes; the three methods have similar median APE for additive AR(1).

\subsection{EI benchmark}
\subsubsection{Within-BM persistence comparison}
\begin{figure}[p]
    \makebox[\textwidth][c]{%
        \begin{minipage}{1.25\textwidth}
            \centering
            \includegraphics[width=\linewidth,height=0.82\textheight,keepaspectratio]{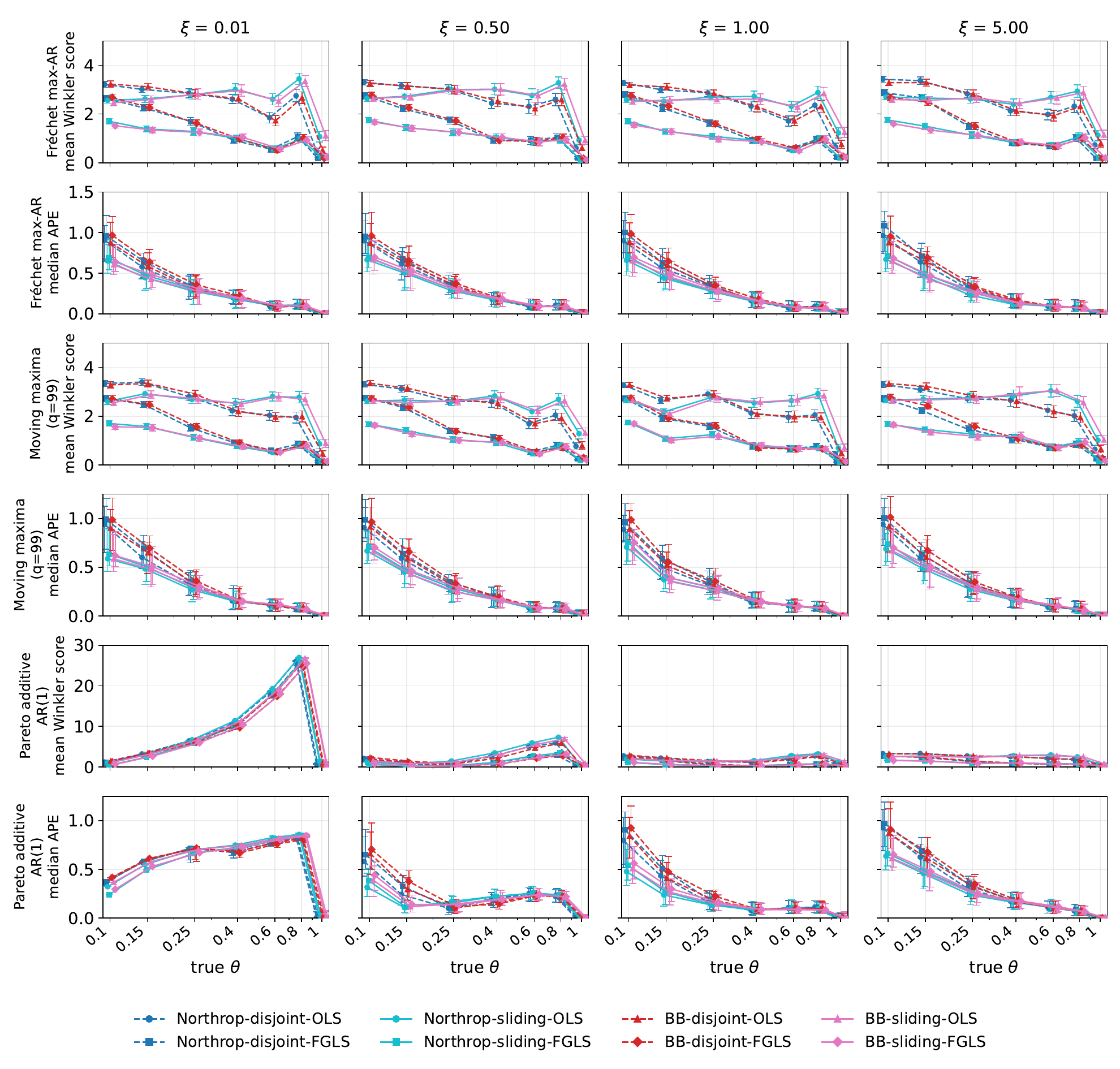}
            \caption{
                Within-block-maxima EI benchmark.
                Panels summarize mean Winkler score for 95\% confidence intervals (\(\pm1\) SE of the mean) and median absolute percentage error (APE, expressed as a fraction; interquartile range) for pooled block-maxima EI variants over the EI benchmark with \(\theta \in \{0.10,0.15,0.25,0.40,0.60,0.80,1.0\}\), \(\xi \in \{0.01,0.50,1.0,5.0\}\), and three dependent heavy-tailed families: Fr\'echet max-AR, moving-maxima \(q=99\), and Pareto additive AR(1).
                Lower values indicate better interval performance or point accuracy.
                The comparison varies EI path construction, block scheme, pooling, and covariance-aware weighting within the persistence branch.
            }
            \label{fig:benchmark-ei-main}
        \end{minipage}}
\end{figure}

\Cref{fig:benchmark-ei-main} summarizes the within-BM EI benchmark across the pooled block-maxima persistence regressions.
BB-sliding-FGLS consistently improves interval performance within the BM framework, yielding lower mean Winkler scores than its pooled OLS counterpart in all \(84\) scenarios. This improvement also holds for the other combinations of EI path and block scheme. Median APE changes little between the sliding-OLS and sliding-FGLS fits.
At \(\xi=1.0\), Winkler scores averaged over the \(\theta\) grid are lower under BB-sliding-FGLS in all three process families (\Cref{tab:benchmark-ei-summary}). For max-AR, the score falls from \(2.38\) under OLS to \(0.91\) under FGLS, while median APE remains \(0.15\); the other families show similarly small changes in median APE.
For BB-sliding, the mean interval width across the full grid increases from \(0.0093\) under OLS to \(0.1352\) under FGLS. FGLS thus alleviates the overly narrow OLS intervals.

Across the within-BM persistence benchmark, BB-sliding-FGLS and Northrop-sliding-FGLS have the two lowest grid-average Winkler scores.
Together, these two sliding-FGLS variants attain the lowest within-BM mean Winkler score in \(54\) of \(84\) scenarios (\(42\) for BB and \(12\) for Northrop); disjoint-FGLS variants attain the minimum in the remaining \(30\).

Both sliding-FGLS variants attain their lowest grid-average Winkler scores at \(\delta=0.15\), with similar scores at the fixed default \(\delta=0.37\); full diagonal shrinkage worsens interval scores and reduces median coverage (\Cref{fig:benchmark-ei-shrinkage} in \ref{app:shrinkage-sensitivity}).

\subsubsection{External baseline comparison}
\begin{figure}[p]
    \makebox[\textwidth][c]{%
        \begin{minipage}{1.25\textwidth}
            \centering
            \includegraphics[width=\linewidth,height=0.82\textheight,keepaspectratio]{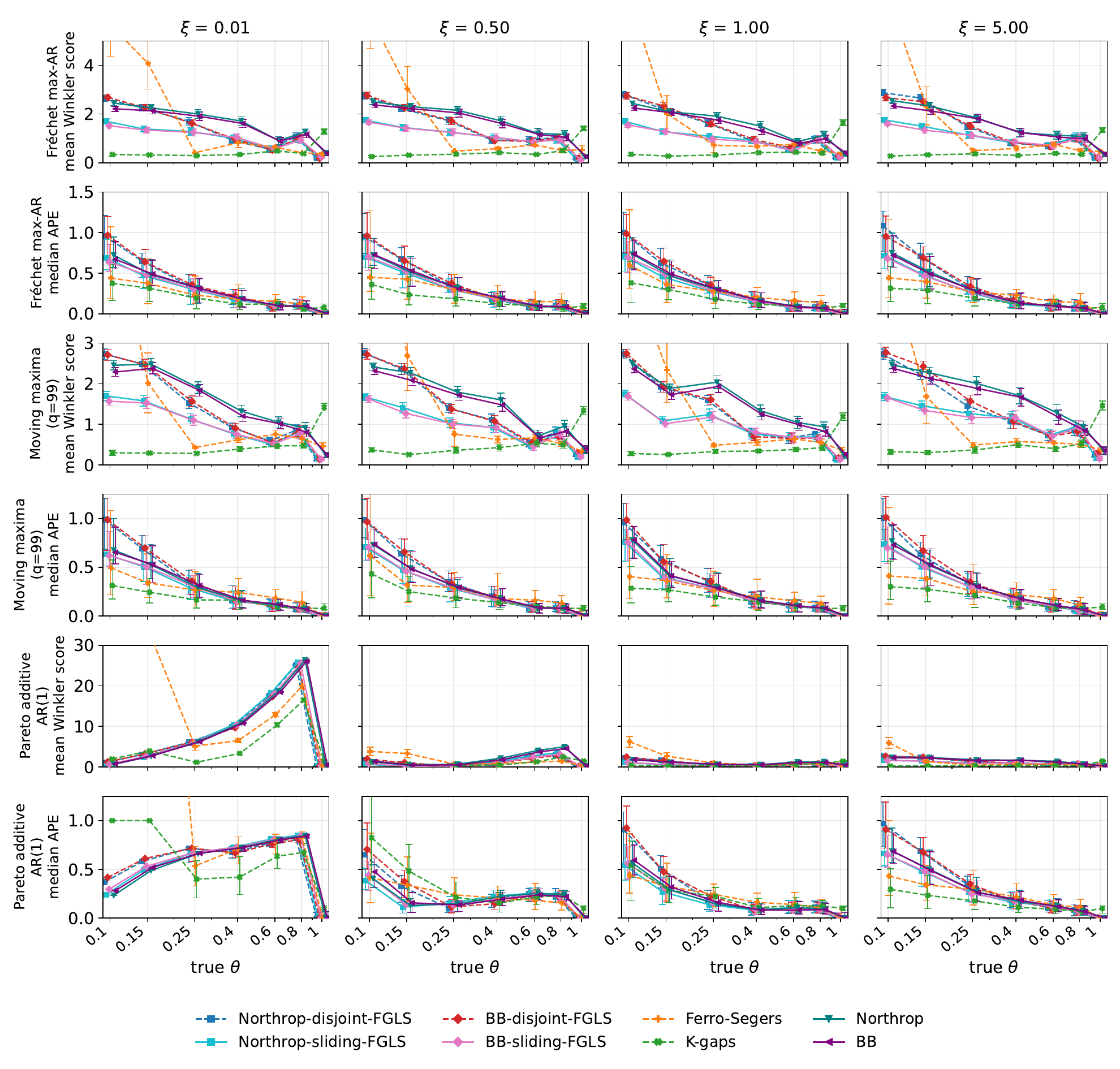}
            \caption{
                EI comparison with external baseline estimators.
                Panels compare selected pooled block-maxima FGLS workflows, using Northrop and Berghaus--B{\"u}cher (BB) EI paths with disjoint or sliding blocks, against Ferro--Segers, K-gaps, and native fixed-\(b\) Northrop and BB estimators over the same EI benchmark.
                Baseline estimators are shown with their native interval constructions.
                Winkler scores use 95\% confidence intervals; bars show \(\pm1\) SE of the mean. APE is a fraction, shown as the median and interquartile range.
                Vertical limits are set from the internal BM methods; external values outside those limits are clipped for display only and remain in the numerical summaries.
                }
            \label{fig:benchmark-ei-targets}
        \end{minipage}}
\end{figure}

Among the eight methods in \Cref{fig:benchmark-ei-targets}, BB-sliding-FGLS and native BB are the only methods that never rank last by mean Winkler score across the \(84\) scenarios.
    BB-sliding-FGLS has lower scores than native BB in \(81\) of the \(84\) scenarios, with grid-average scores of \(1.61\) and \(2.08\), respectively.
    K-gaps ranks last in all \(12\) scenarios with \(\theta=1\).

After these baselines are added, K-gaps has the lowest mean Winkler score in \(59\) of the \(84\) scenarios and the lowest grid-average Winkler score, \(1.00\), compared with \(1.61\) for BB-sliding-FGLS and \(1.65\) for Northrop-sliding-FGLS.
\ref{app:benchmark-baselines} summarizes the threshold-based and native fixed-\(b\) baseline models used in this comparison.

The limited threshold grid used for K-gaps and Ferro--Segers may not be equally suitable across dependence structures.
    Both estimators select among the empirical 0.90 and 0.95 quantile thresholds under pre-specified rules; K-gaps also considers the run parameter \(K\in\{1,2\}\), measured in observation steps.
    The comparison is therefore conditional on these candidate sets and selection rules.

Averaged over the \(\theta\) grid, BB-sliding-FGLS has lower Winkler scores than K-gaps for Pareto additive AR(1) at both \(\xi=0.5\) and \(1.0\) (\Cref{tab:benchmark-ei-summary}). On the \(\xi=1.0\) slice, K-gaps has lower scores for max-AR and moving-maxima processes, while BB-sliding-FGLS has lower grid-average Winkler scores than Ferro--Segers and the native fixed-\(b\) BM estimators in all three process families.

For the Pareto additive AR(1) family at \(\xi=0.01\), method-specific median empirical coverage across the seven dependence settings ranges from \(0\%\) to \(3\%\), compared with the nominal \(95\%\).

\section{Applications}\label{sec:applications}
\subsection{Data sources and application design}
The applications examine streamflow and insured flood losses in Texas and Florida.
Daily mean discharge records are obtained from the U.S. Geological Survey (USGS), and daily building-claim totals from National Flood Insurance Program (NFIP) records.
The paired applications provide a descriptive comparison of hydrologic hazards and insured losses within each state \citep{czajkowski_economic_2016,nelson_mercer_pluvial_2025}.
Texas and Florida are retained because both domains can be constructed from public data sources under comparable preprocessing rules.

These applications illustrate how to interpret severity and extremal clustering estimates, together with their uncertainty, for streamflow and insured flood losses.
Throughout this section, design-life levels are defined on the relevant observation clock; see \Cref{sec:design-life-interpretation}.
For streamflow, the clock is the calendar-day record at the selected gauge.
For NFIP severity, it is the active-day record.
\ref{app:application-uncertainty} reports window-selection sensitivity and USGS gauge-screening results.
The application fits use the archived provider data, fixed covariance shrinkage \(\delta=0.37\), and the adaptive bootstrap rule in \ref{app:implementation-defaults}. All parameter fits meet the finite-bootstrap MCSE criterion. Case-specific replicate counts are reported in \ref{app:application-computation}.

\begin{table}[htbp]
\centering
\scriptsize
\setlength{\tabcolsep}{3pt}
\renewcommand{\arraystretch}{1.05}
\caption{Cross-application summary for the four application cases. Parameter entries report the severity-side estimate \(\widehat\xi\) and the persistence-side estimate \(\widehat\theta\) with 95 percent confidence intervals. Mean cluster size is reported as the implied point summary \(1/\widehat\theta\), to keep the main-text table compact; its confidence interval is inherited by reciprocal transformation of the reported \(\widehat\theta\) interval. The 10- and 50-year columns report the median design-life level with its conditional 95 percent confidence interval from the selected fit, rescaled for readability: streamflow entries are in \(10^3\,\mathrm{ft}^3\,\mathrm{s}^{-1}\), and NFIP entries are in \(10^6\) 2025 U.S. dollars. Streamflow uses the calendar-day clock for both severity and persistence; NFIP uses the active-day clock for severity and the calendar-day clock for persistence.}
\label{tab:application-summary}
\begin{tabularx}{\textwidth}{>{\raggedright\arraybackslash}p{0.22\textwidth}>{\centering\arraybackslash}X>{\centering\arraybackslash}X>{\centering\arraybackslash}X>{\centering\arraybackslash}X>{\centering\arraybackslash}X}
\toprule
Application & $\widehat\xi$ & $\widehat\theta$ & \shortstack[c]{Mean \\ cluster size} & \shortstack[c]{10y \\ design-life \\ level} & \shortstack[c]{50y \\ design-life \\ level} \\
\midrule
Texas streamflow & 0.64 [0.59, 0.70] & 0.05 [0.04, 0.05] & 20.50 & 232 [193, 280] & 655 [503, 854] \\
Florida streamflow & 0.33 [0.06, 0.59] & 0.06 [0.05, 0.06] & 18.14 & 38.5 [9.64, 154] & 65.0 [10.6, 398] \\
Texas NFIP claims & 1.40 [0.72, 2.07] & 0.31 [0.28, 0.35] & 3.21 & 478 [24.1, 9,466] & 4,509 [76.8, 264,684] \\
Florida NFIP claims & 1.38 [1.05, 1.72] & 0.31 [0.27, 0.35] & 3.23 & 224 [47.3, 1,058] & 2,072 [257, 16,688] \\
\bottomrule
\end{tabularx}
\end{table}

\subsection{Streamflow: daily mean discharge}
The streamflow applications use daily mean discharge from USGS gauges 08066500 (Trinity River at Romayor, Texas) and 02366500 (Choctawhatchee River near Bruce, Florida).
    The Texas record spans 1 May 1924 to 31 December 2025 (37,135 observations; 101.7 years), and the Florida record spans 1 June 1984 to 31 December 2025 (15,189 observations; 41.6 years).
    Both branches use the same continuous calendar-day record at each gauge.
For both sites, repeated dates and invalid discharge values are removed, after which the analysis uses the longest gap-free daily suffix ending at the fixed cutoff of 31 December 2025; missing days are not imputed and the remaining dates are not compressed.
    The conventional generalized extreme value (GEV) scale check uses maxima of daily mean discharge from these same records, retaining January--December years with at least 97\% valid daily observations.
    The retained years are 1925--2025 for Texas (101 maxima) and 1985--2025 for Florida (41 maxima); all have complete daily coverage.
Design-life levels are reported in cubic feet per second (cfs; \(\mathrm{ft}^3\,\mathrm{s}^{-1}\)); EI is dimensionless and describes clustering of daily observations.
The block-size diagnostics assess the log-quantile and transformed EI paths; they do not test temporal stationarity of the discharge record.

In the severity branch, the estimated extreme value index \(\widehat\xi\) and the derived design-life curves summarize how the \(T\)-year horizon maximum of daily discharge grows as the design life increases \citep{rootzen_katz_design_life_2013,salas_obeysekera_2014,read_vogel_2015}.
In the persistence branch, \(\widehat\theta\) summarizes temporal clustering of extreme daily flows, providing a statistical descriptor of high-flow persistence \citep{northrop_efficient_2015}.
    The reciprocal \(1/\widehat\theta\) estimates the limiting mean number of upper-tail exceedances per cluster \citep{ferro_segers_2003}.

\begin{figure}[p]
    \centering
    \includegraphics[width=\textwidth,height=0.38\textheight,keepaspectratio]{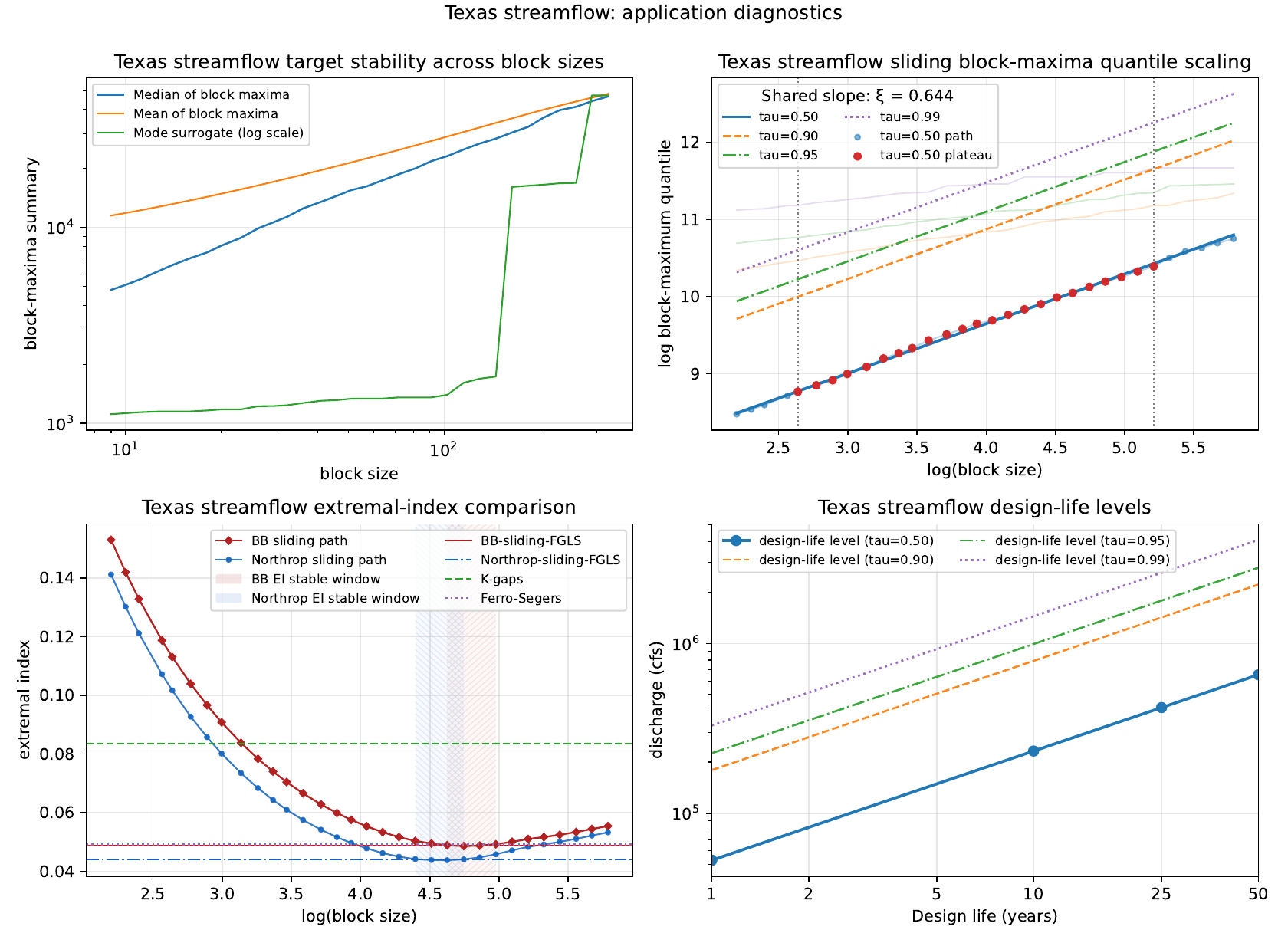}\par\medskip
    \includegraphics[width=\textwidth,height=0.38\textheight,keepaspectratio]{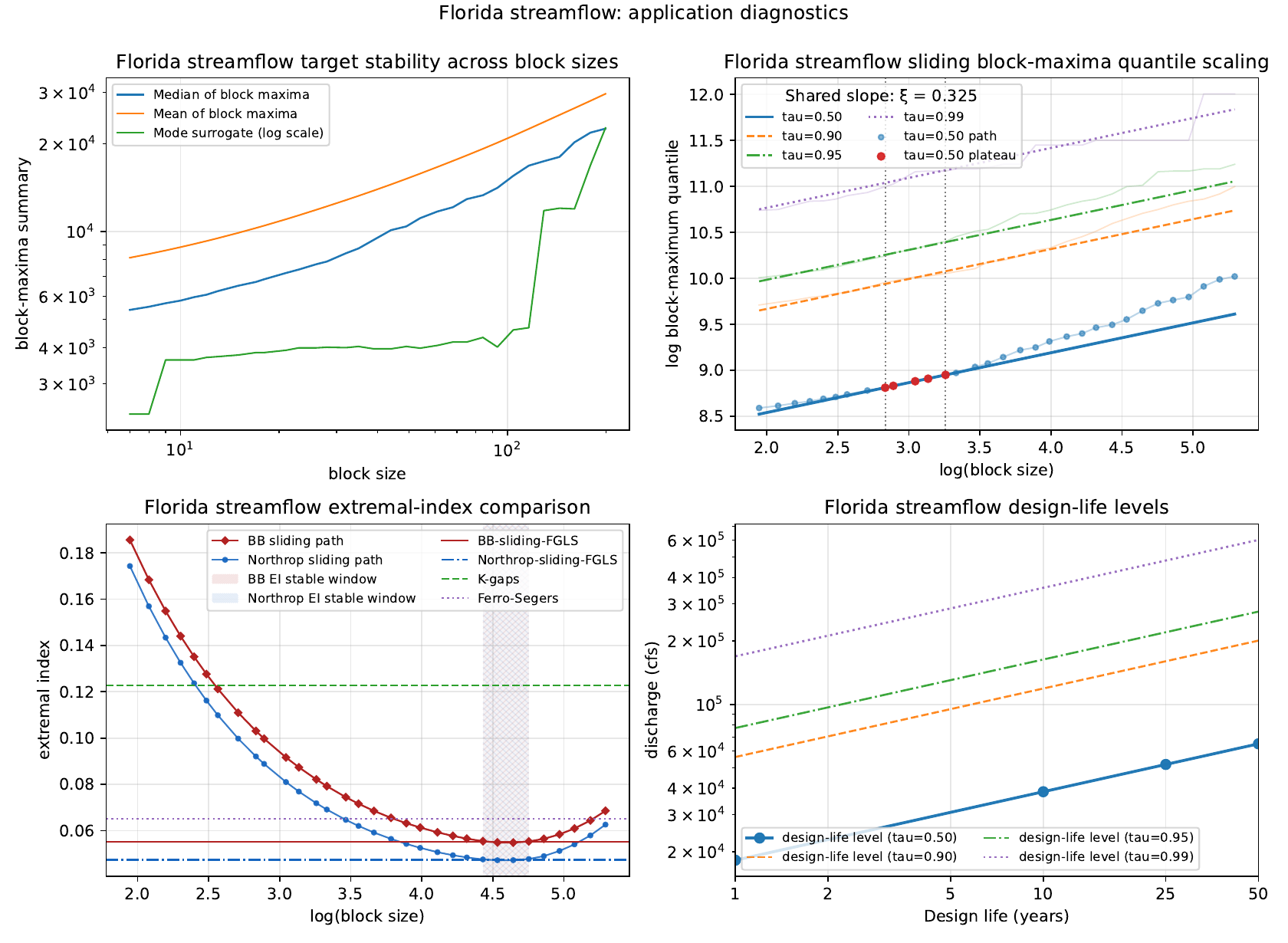}
    \caption{
        Streamflow applications in Texas (top) and Florida (bottom).
        For each state, panels show block-maximum summaries, shared-slope quantile fits, EI paths and estimates, and design-life levels, reading from upper left to lower right.
        EI shading marks the BB and Northrop EI stable windows; horizontal lines show their pooled FGLS estimates and the K-gaps and Ferro--Segers estimates.
            Design-life curves use \(\tau=0.50,0.90,0.95,0.99\).
            Both branches use calendar days, with discharge in \(\mathrm{ft}^3\,\mathrm{s}^{-1}\).
    }
    \label{fig:application-streamflow}
\end{figure}

The selected block-size windows give a larger EVI estimate for Texas than for Florida, with \(\widehat\xi \approx 0.64\) versus \(\widehat\xi \approx 0.33\) (\Cref{tab:application-summary,fig:application-streamflow}).
Under the fitted stationary model, the median 10- and 50-year design-life levels are larger for Texas, and the fitted scaling relation implies faster growth of high-discharge quantiles as the design life increases.
These design-life levels are obtained by extrapolating the fitted scaling relation on the calendar-day clock; they should not be read as calendar-year-maxima return levels or as replacements for Bulletin~17C or GEV flood-frequency estimates.
Uncertainty in extreme-level estimates can propagate into subsequent inundation assessments \citep{fang_extreme_2021}.

The relatively large finite-block EVI estimate at Romayor may reflect seasonal flow variability over the selected 14--183-day block-size range.
    The gauge lies downstream of Lake Livingston in the regulated lower Trinity \citep{wellmeyer_impoundment_2005}, where channel--floodplain exchanges influence high-flow conveyance \citep{phillips_slattery_downstream_2007}.
    As blocks lengthen from weeks to months, they are more likely to encompass seasonal high-flow episodes, which can contribute to rapid growth of the median block maximum.
    This provides a plausible interpretation of the observed scaling, although the contributions of seasonality, flood-generating mechanisms, and regulation have not been isolated.

For context, \citet{tyralis_shape_2019} reported a median GEV shape estimate of 0.19 for annual maxima of daily streamflow across 591 minimally disturbed U.S. basins.
    The authors document the exclusion of basins with estimated GEV shape parameters \(\ge1\), for which the fitted distribution has no finite mean \citep{tyralis_shape_preprint_2018}.
    Florida's EVI estimate of 0.33 has a conditional 95\% interval of [0.06, 0.59], which includes this cross-basin reference value.
    Under stationary regular variation and a positive extremal index, both approaches target the same asymptotic EVI.
    For Texas, the finite-block scaling fit (\(\widehat\xi=0.64\), conditional 95\% interval [0.59, 0.70]) yields substantially higher extrapolated discharge levels than the annual-maxima GEV fit (\Cref{tab:application-streamflow-gev-check}).
    We interpret the Texas EVI estimate over the selected 14--183-day block-size range and present the 50-year design-life level as an illustration.

The estimated extremal indices imply limiting mean cluster sizes of approximately \(20.5\) upper-tail exceedances of daily discharge in Texas and \(18.1\) in Florida (\Cref{tab:application-summary}).

The selected severity and EI windows need not coincide, as the branches use different stability criteria (\Cref{fig:application-streamflow}).

\subsection{NFIP claims: daily building-claim totals}
The NFIP applications assess severity and persistence in daily building-claim totals.
For each state, we construct daily building-claim totals by summing NFIP building-claim payments from OpenFEMA statewide by date of loss over 1 January 1978 to 31 December 2025 (48 calendar years).
Monetary values are deflated by loss month to 2025 U.S. dollars (USD) using the monthly, not-seasonally-adjusted Consumer Price Index for All Urban Consumers (CPI-U) for the U.S. city average \citep{bls_cpi_u_historical}. The base value and treatment of the missing October 2025 observation are documented in \ref{app:application-computation}.

The resulting daily building-claim series is analyzed on two observation clocks.
Severity is fitted on active days, whereas persistence is fitted on the calendar-day building-claim series covering all 17,532 days in the 1978--2025 analysis period.
    A zero denotes a zero total of recorded building-claim payments for losses on that date.
    An active day is any date with a strictly positive daily building-claim total; this definition does not require the total to be extreme.

A severity block of \(b\) observations contains \(b\) active days, which need not be consecutive in calendar time; an EI block of size \(b\) spans \(b\) consecutive calendar days.
    We map a \(T\)-year design life to \(b_T=\lceil\widehat\lambda_{\mathrm{active}}T\rceil\), where \(\widehat\lambda_{\mathrm{active}}\) is the mean number of active days per year over the retained analysis period.
    This conversion assumes that the historical active-day rate remains applicable; variability in the future number of active days is not modeled.

For Texas and Florida, the 50-year conversion gives \(b_{50}=6{,}049\) and \(5{,}667\) active days, respectively.
    Under the stationary working model, the estimated medians of the maximum daily building-claim total over these blocks are \(\$4.51\) billion and \(\$2.07\) billion in 2025 U.S. dollars.
    The corresponding conditional 95\% confidence intervals are \([\$0.077,\$264.68]\) billion and \([\$0.257,\$16.69]\) billion.
    Computed using \eqref{eq:design_life_ci}, these intervals quantify estimation uncertainty in \(D_{0.5}(50)\), not variability of future maxima.
    Their width reflects joint parameter uncertainty amplified by extrapolation beyond the fitted block-size windows (\Cref{sec:discussion}).

\begin{figure}[p]
    \centering
    \includegraphics[width=\textwidth,height=0.38\textheight,keepaspectratio]{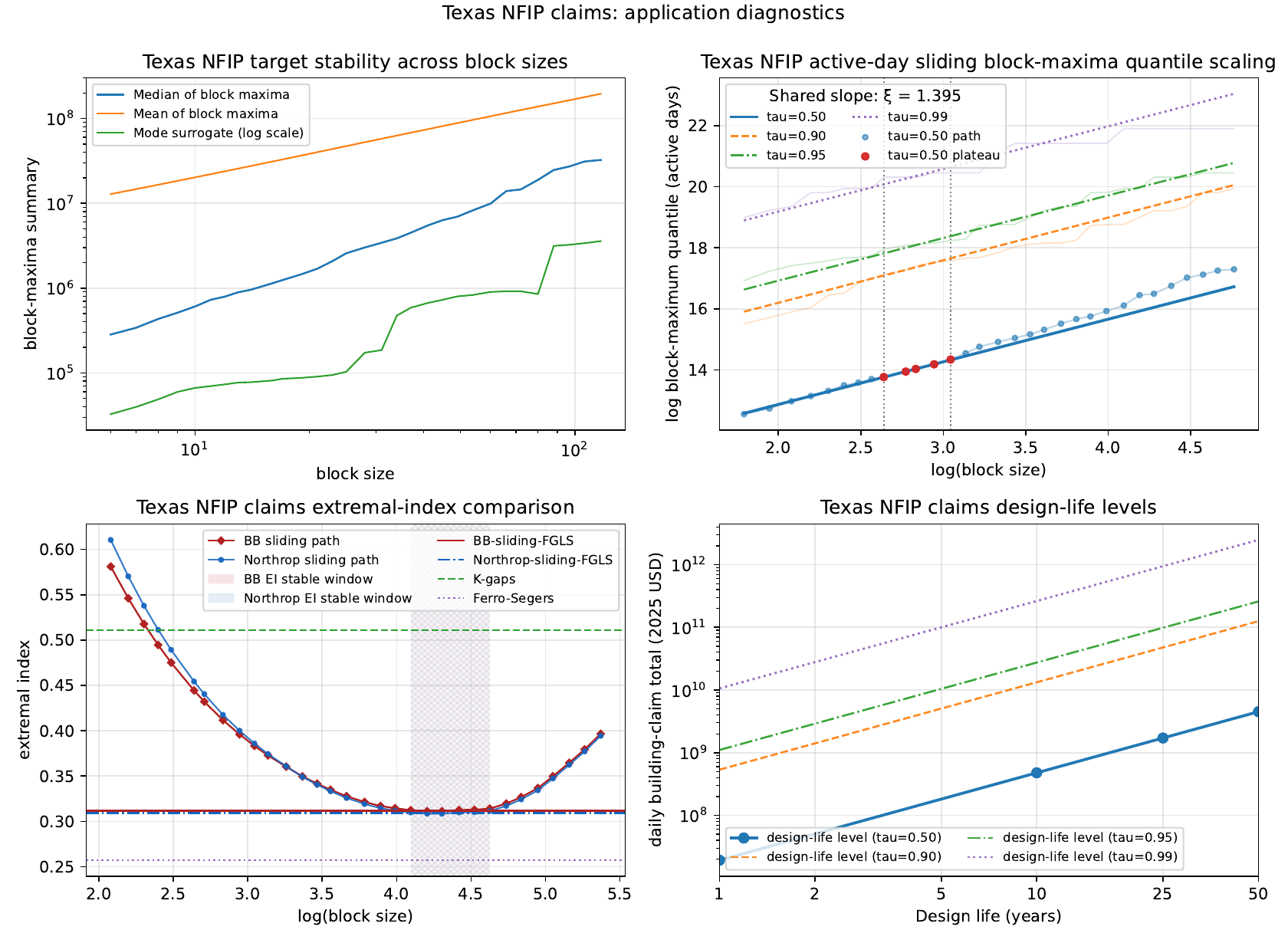}\par\medskip
    \includegraphics[width=\textwidth,height=0.38\textheight,keepaspectratio]{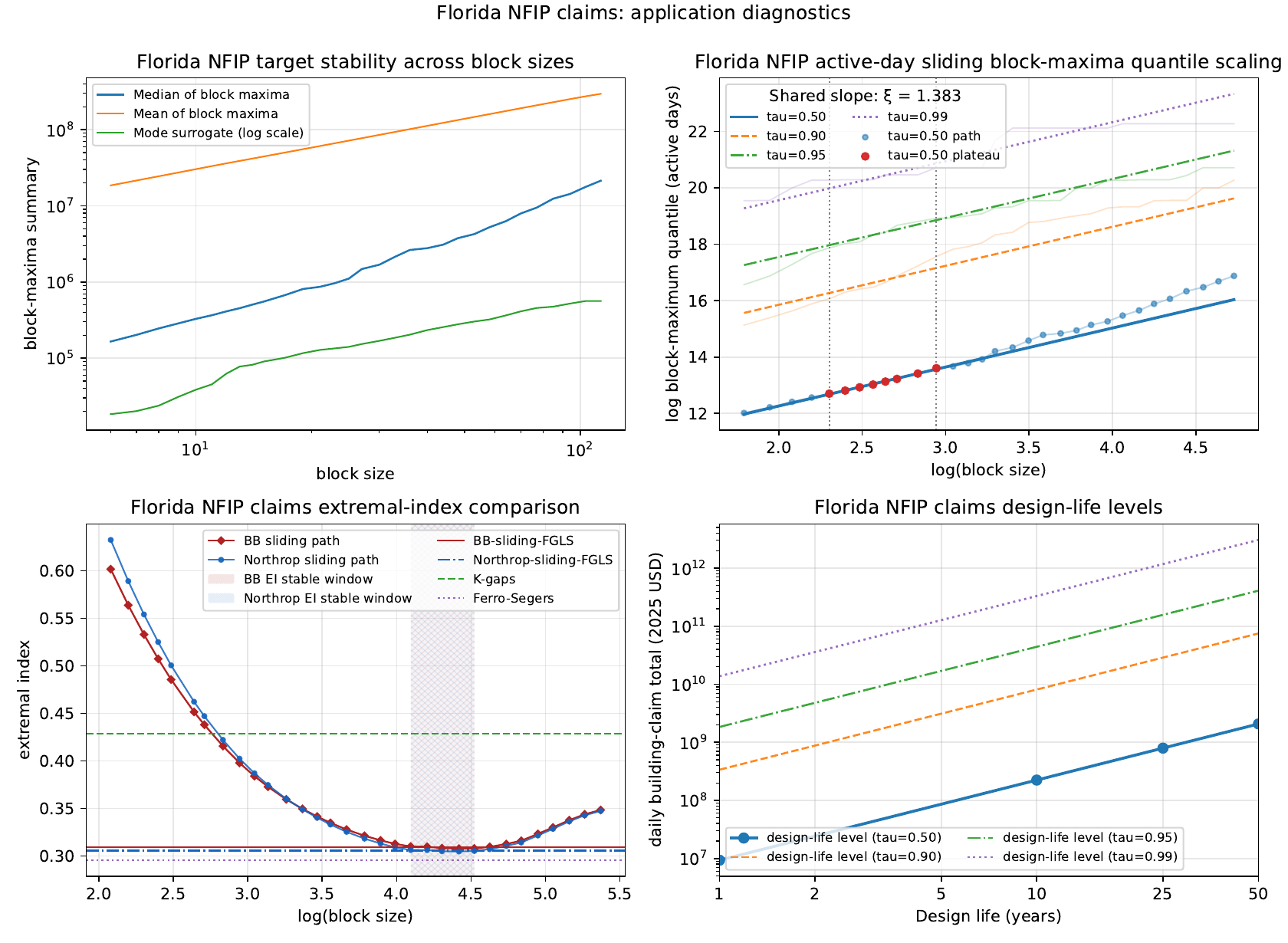}
    \caption{
        NFIP building-claim applications in Texas (top) and Florida (bottom).
        For each state, panels show block-maximum summaries, shared-slope quantile fits, EI paths and estimates, and design-life levels, reading from upper left to lower right.
        EI shading marks the BB and Northrop EI stable windows; horizontal lines show their pooled FGLS estimates and the K-gaps and Ferro--Segers estimates.
            Design-life curves use \(\tau=0.50,0.90,0.95,0.99\), with years converted using the historical active-day rate.
            Severity uses active days and EI uses calendar days; building-claim totals are grouped by date of loss and expressed in 2025 USD.
    }
    \label{fig:application-nfip}
\end{figure}

Texas NFIP claims yield \(\widehat\xi \approx 1.40\), and Florida NFIP claims yield \(\widehat\xi \approx 1.38\) (\Cref{tab:application-summary,fig:application-nfip}).
These estimates are larger than the corresponding streamflow estimates, but their interpretation is different: they describe the tail of daily building-claim totals on the active-day clock.
These design-life levels support exploratory stress assessment.
If interpreted as asymptotic EVI values, the NFIP estimates above one would imply no finite mean under an unbounded regularly varying model \citep{arnold_pareto_2014}.
    Fixed-probability quantiles remain finite; the reported design-life levels describe quantiles of the maximum daily building-claim total, not expected annual losses.

Both NFIP cases have \(\widehat\theta \approx 0.31\), corresponding to limiting mean cluster sizes of about \(3.21\) upper-tail exceedances in Texas and \(3.23\) in Florida (\Cref{tab:application-summary}).
These counts refer to extreme daily building-claim totals, not all active days or elapsed cluster durations.

\subsection{Cross-application synthesis}
The finite-block EVI estimates are \(0.33\)--\(0.64\) for streamflow and \(1.38\)--\(1.40\) for NFIP daily building-claim totals on active days; the corresponding calendar-time EI estimates are about \(0.05\) and \(0.31\), respectively.
    In these records, streamflow shows stronger extremal clustering, while the NFIP severity fits imply faster growth of maximum daily building-claim totals on the active-day clock as the design life increases.

The two streamflow records have comparable calendar-time EI estimates but different EVI estimates.
    This pattern illustrates the distinction between marginal tail behaviour and temporal extremal dependence: strictly increasing marginal transformations preserve the EI while potentially changing tail heaviness \citep{papastathopoulos_markov_2017}.
    Routing and catchment storage can sustain clusters of high flows, providing a plausible hydrological context for the low EI estimates \citep{darcy_tawn_river_flows_2025}.

For NFIP, active-day EVI and calendar-time EI estimates are similar between states within the retained records and fitted block-size ranges.
    Broader NFIP analyses document heavy-tailed loss margins \citep{li_factor_copula_2026} and substantial concentration of losses within major disasters \citep{nayak_hyperclustering_2025}.
    Similar EVI and EI estimates can coexist with different loss magnitudes and regional dependence structures.

\section{Discussion and conclusions}\label{sec:discussion}

The block-maxima workflow combines data-adaptive block-size selection with covariance-aware estimation of the EVI and EI for serially dependent hydrologic and flood-impact records, and derives design-life levels from the fitted severity scaling.

The stationary synthetic benchmark with \(N=365\) observations per record indicates that the clearest gains are in interval performance rather than point accuracy.
For EVI, median-sliding-FGLS achieves the lowest grid-average Winkler score and is the only method that never ranks last by mean Winkler score in the \(84\)-scenario external comparison.
    For EI, BB-sliding-FGLS and native BB are the only methods that never rank last by mean Winkler score in the corresponding comparison (\Cref{fig:benchmark-evi-targets,fig:benchmark-ei-targets}).
    BB-sliding-FGLS has lower mean Winkler scores than native BB in \(81\) scenarios and than its pooled OLS counterpart in all \(84\).

Long-horizon interpretation requires substantial extrapolation in all four applications.
    For Florida streamflow, the fitted window spans 17--26 calendar days, whereas the 50-year target \(b_{50}=18{,}263\) is about 702 times the upper fitted block size.
    Each NFIP scaling fit uses block maxima formed from all active-day observations in the 48-year observation period.
    The selected block sizes are 14--21 active days in Texas and 10--19 in Florida; the 50-year targets of 6,049 and 5,667 active days are respectively 288 and 298 times the largest fitted block sizes (\Cref{tab:application-extrapolation}).

The wide NFIP intervals reflect joint parameter uncertainty amplified by extrapolation.
    Both EVI estimates are near \(1.4\), but the Texas estimate has a wider confidence interval.
    The interval width in \eqref{eq:design_life_ci} depends on both coefficient variances, their covariance, and the target block size; exponentiation produces asymmetric dollar-scale intervals.
    At a 300-fold increase in block size, increasing \(\xi\) by \(0.1\) multiplies the extrapolated level by \(300^{0.1}\approx1.77\), holding the fitted median at the reference block size fixed.
    Sensitivity to the largest losses and tail-fitting choices is a recognized limitation of insurance tail inference \citep{mcneil_loss_tails_1997}.

Further work should test scaling at larger block sizes and evaluate coverage with window-selection uncertainty included.
    For NFIP, additional comparable extreme-loss records and explicit modeling of changes in exposure and policy coverage could strengthen inference.

The NFIP curves support exploratory stress assessment of the largest daily building-claim total over a specified number of active days.
    Under the fitted model and historical active-day-rate conversion, \(\tau=0.95\) corresponds to a 5\% probability that this maximum exceeds the fitted level.
    Cumulative losses and financing needs require additional modeling; totals grouped by date of loss do not describe cash-payment timing.

The workflow supplements standard flood-frequency analysis with summaries of severity scaling and extremal clustering.
In the United States, Bulletin~17C recommends fitting a Log-Pearson Type~III distribution to annual peak discharges, with regional skew and historical information incorporated when available \citep{england_bulletin_17c_2019}.
Annual-maxima GEV models and peaks-over-threshold analyses with generalized Pareto tails are also widely used in hydrologic extremes \citep{davison_smith_1990,coles_introduction_2001}.
The GEV and generalized Pareto families accommodate positive, zero, and negative shape parameters; the present severity scaling assumes a regularly varying upper tail with \(\xi>0\).
    Peaks-over-threshold methods can use several extremes per year, with threshold choice and clustering handled explicitly \citep{katz_statistics_2002}.
    GEV and generalized Pareto models can be fitted directly to block maxima and threshold exceedances, respectively, without prior mean--variance or autoregressive moving-average (ARMA) filtering.

The autoregressive models generate benchmark series; the workflow does not require fitting them to the observations.
    Stationary ARMA(1,1) processes with regularly varying innovations may satisfy the stated assumptions under suitable coefficient conditions \citep{rootzen_moving_average_1986}; the workflow's finite-sample performance for this class remains to be evaluated.

Regional flood-frequency analysis pools information across hydrologically comparable sites to support estimation at a target site \citep{hosking_wallis_regional_1997}.
    Under the stated assumptions, the present workflow combines information across block sizes within a single record for severity and clustering assessment.
    Conventional annual-peak design and regional or ungauged-site estimation should follow the corresponding established flood-frequency methods \citep{england_bulletin_17c_2019,hosking_wallis_regional_1997}.
    Comparisons with standard frequency estimates require aligned periods, observation scales (e.g., daily mean versus instantaneous peak discharge), and probability definitions.

A mass at zero is compatible with a regularly varying upper tail.
    The workflow accommodates independent observations, represented in the benchmark, as well as sequences satisfying the stated dependence conditions.
    In rainfall models, a Markov wet--dry occurrence process can coexist with conditionally independent positive amounts \citep{furrer_katz_precipitation_2008}.
    The extremal index quantifies clustering of extreme rainfall amounts, which is distinct from wet-spell persistence.
    Application requires the tail and dependence assumptions and enough positive block summaries; sparse records may need an occurrence--amount model.
    Inter-arrival times or spell durations provide alternative descriptions of rainfall occurrence \citep{baiamonte_rainfall_spells_2024}; applying the severity branch to these variables would target duration extremes under its tail and dependence assumptions.
    Removing dry days changes the observation clock.
    The NFIP example does not validate performance for zero-heavy rainfall.

The applications use a stationary working model for each retained record.
    Climate variability and change, land-use change, and river regulation may affect streamflow extremes or clustering.
    CPI-U adjustment of NFIP building-claim payments addresses inflation without controlling for changes in exposure, coverage, or reporting.
    Such changes are not modeled explicitly or included in the reported uncertainty.
    The estimates are conditional summaries of the retained records; records with clear trends or regime shifts should not be pooled under this stationary workflow for direct future-design extrapolation.

Physical evidence and temporal diagnostics can guide the choice of approximately stationary periods \citep{ryberg_change_points_2020,slater_nonstationary_2021}.
    The workflow may be applied within each period under the stated assumptions, provided sufficient extreme-event information is retained.
    Inference following data-driven segmentation, including uncertainty in period boundaries, requires separate validation.

Seasonal or covariate-dependent extensions would require models of temporal variation \citep{chavez_demoulin_davison_2005}, with future covariate values or scenarios specified for projection.
    \citet{barbaux_integrating_2025} estimate nonstationary design-life levels from Bayesian predictive distributions that incorporate both stochastic and estimation uncertainty over the design period.
    Spatial or multivariate extensions would require models of joint tail dependence \citep{huser_advances_2022}.

\section*{Declarations}

\subsection*{Ethics approval} Not applicable.

\subsection*{Data and code availability}
The UniBM software used for computation, benchmarking, and visualization is available at
\href{https://github.com/TY-Cheng/UniBM}{https://github.com/TY-Cheng/UniBM}
under the MIT License.
The repository contains the workflow code, fixed-cutoff public-provider extracts, source-metadata records, and 
derived analysis outputs used to generate the reported figures and tables.
The submission package contains the manuscript source and the derived figures and tables needed to compile the article.
The tracked provider extracts are bounded by the declared analysis cutoff of 31 December 2025, so the reported analysis can be rebuilt without downloading live records.
Instructions for regenerating the reported figures and tables are provided in the repository \texttt{README}.
The reported analyses use U.S. Geological Survey streamflow records, OpenFEMA NFIP claims data, and the U.S. Bureau of Labor Statistics CPI-U series.
External source records remain subject to the terms, access conditions, and availability of the original providers.

\subsection*{Author contributions}
T.C. contributed conceptualization, methodology, software, formal analysis, visualization, and writing of the original draft.
X.P. contributed data interpretation, validation, and writing--review and editing.
A.C. contributed methodology, validation, and writing--review and editing.
X.H. contributed hydrologic framing, supervision, and writing--review and editing.
K.C. contributed conceptualization, supervision, funding acquisition, and writing--review and editing.
All authors reviewed and approved the manuscript.

\subsection*{Declaration of competing interest}
Xiaogang He serves as an Associate Editor of \textit{Journal of Hydrology}.
Given this role, he had no involvement in the peer review of this article and had no access to information regarding its peer review.
Full responsibility for the editorial process for this article was delegated to another journal editor.
The other authors declare no known competing financial interests or personal relationships that could have appeared to influence the work reported in this paper.

\subsection*{Funding}
This research was partly supported by the Singapore Ministry of Education academic research grant R-144-000-457-733 on Outbreak Resilience, Forecasting, and Early Warning.
The funder had no role in study design, data collection and analysis, decision to publish, or preparation of the manuscript.

\bibliographystyle{elsarticle-harv}
\bibliography{1_references}

\newpage
\appendix
\counterwithin*{figure}{section}
\counterwithin*{table}{section}

\section{Methodological details}

\subsection{Implementation defaults}\label{app:implementation-defaults}

\Cref{tab:implementation-defaults} reports the pre-specified implementation defaults used in all benchmark and application analyses.
They are treated as fixed computational settings, not as data-optimized tuning choices.

\begin{table}[htbp]
    \centering
    \scriptsize
    \setlength{\tabcolsep}{4pt}
    \caption{Pre-specified implementation defaults for the severity and persistence branches.}
    \label{tab:implementation-defaults}
    \begin{tabular}{p{0.48\textwidth}p{0.32\textwidth}}
        \toprule
        Setting (fixed value)                                                                                                                 & Purpose                                                                                                 \\
        \midrule
        Severity covariance bootstrap replicates (adaptive \(R\), cap \(1024\))                                    & super-block bootstrap covariance for the log-quantile path               \\
        Persistence covariance bootstrap replicates (adaptive \(R\), cap \(1024\))                                 & circular block-bootstrap covariance for the transformed EI path           \\
        Super-block length (\(s^\star=\max\{4b_{\max,\mathrm{grid}},\sqrt{N}\}\)) & dependence-preserving resampling for severity            \\
        Covariance shrinkage (\(\delta=0.37\))                                                               & diagonal shrinkage before FGLS inversion                                                                \\
        Ridge term (\(r_{\mathrm{ridge}}=\max\{(\mathrm{tr}(\widehat\Sigma)/p)10^{-8},10^{-12}\}\))                & numerical stabilization of covariance inversion                                                         \\
        EVI curvature penalty (\(\lambda_{\mathrm{curv}}=2.0\))                                       & plateau selection                                                \\
        EI stable window minimum points (\(4\))                                                                & minimum number of points in the EI stable window \\
        EI trim fraction (\(0.15\))                                                                                                           & local stabilization of the EI path                                                                      \\
        EI roughness penalty (\(0.75\))                                                                                                       & EI stable window scoring                                                 \\
        EI curvature penalty (\(0.5\))                                                                                                        & EI stable window scoring                                                 \\
        \bottomrule
    \end{tabular}
\end{table}

Here \(b_{\max,\mathrm{grid}}\) is the largest candidate block size used in the full retained log-quantile path, and \(p\) is the dimension of the covariance matrix being regularized.
The operator \(\mathrm{tr}\) denotes the matrix trace.
Adaptive bootstrap replication uses checkpoints \(R\in\{128,256,512,768,1024\}\).
    At each checkpoint, a delete-group jackknife over whole bootstrap rows uses eight groups and two random partitions to estimate finite-bootstrap MCSE.
    The rule stops at the first checkpoint at which every monitored target has estimated MCSE no greater than 10\% of its reference statistical SE, or at \(R=1024\).
    The monitored targets are \(\widehat\xi\) and its interval endpoints for EVI, and \(\widehat\theta\), its interval endpoints, the unconstrained pooled \(z\) estimate, and its interval endpoints for EI.
    The reference scales are \(\operatorname{SE}(\widehat\xi)\) for all EVI targets, \(\widehat\theta\,\widehat s_\mu\) for the three \(\theta\)-scale targets, and \(\widehat s_\mu\) for the three unconstrained \(z\)-scale targets.
    If the cap is reached without meeting the criterion, the fit is retained and the unmet precision criterion is recorded.

\subsection{Severity-side super-block bootstrap construction}\label{app:severity-bootstrap-backbone}
The series is partitioned into contiguous, non-overlapping super-blocks of equal length, exceeding the largest block size used for the fit; an incomplete final super-block is omitted.
    For the sliding-block variant at block size \(b\), the first \(b-1\) observations of each super-block are appended to its end before computing sliding maxima, following the circular construction of \citet{bucher_staud_bootstrap_2026}.
    For the disjoint-block variant, maxima are computed from complete non-overlapping blocks within each super-block.
    Each bootstrap replicate draws as many super-block indices as there are retained super-blocks, with replacement, and uses the same indices for all block sizes.
    At each block size, the precomputed maxima from the sampled super-blocks are pooled, and the selected summary (median, mean, or mode) is recomputed.
The mode baseline uses a smoothed kernel-density surrogate on the log block-maxima scale \citep{cheng_mean_1995,aliyari_ghassabeh_convergence_2013}.

Median, mean, and mode targets use the same seeded sequence of super-block draws to estimate their respective covariance matrices.
    Adaptive stopping is applied separately to each target, so the covariance estimates may use different numbers of draws from this common sequence.

\subsection{Workflow algorithms}\label{app:algorithms}
\Cref{alg:severity-workflow,alg:persistence-workflow}
summarize the severity and persistence implementations used in the benchmark and applications.
Each algorithm uses the input series and block-size grid for the specified observation clock and is run separately for each estimator configuration.
    The severity algorithm presents the median target; the mean and mode benchmark variants replace the median summary in the parameter-estimation steps.
    Both FGLS variants hold the selected original-data window fixed and use the regularization and adaptive-bootstrap defaults in \ref{app:implementation-defaults}.
    The severity resampling construction is described in \ref{app:severity-bootstrap-backbone}.

\begin{algorithm}[htbp]
    \caption{Severity workflow for EVI and design-life levels}
    \label{alg:severity-workflow}
    
        \begin{algorithmic}[1]
            \Require Series \(X_1,\ldots,X_N\), block-size grid \(\{b_k\}_{k=1}^{K}\), block scheme (sliding or disjoint), and fitting method (OLS or FGLS)
            \Require For application outputs: observation rate \(\lambda\) and requested design lives
            \For{each block size \(b_k\)}
            \State Extract block maxima using the specified block scheme
            \State Compute the empirical median \(\widehat Q_{0.5}(M_{b_k})\)
            \EndFor
            \State Retain finite positive summaries and form the log-quantile path
            \State Select the plateau minimizing \eqref{eq:plateau_score}
            \If{FGLS is selected}
            \State Estimate the log-quantile covariance by adaptive super-block bootstrap
            \State Regularize the selected covariance and fit by FGLS using \Cref{eq:covariance_regularization,eq:fgls_est}
            \Else
            \State Fit the scaling relation by OLS on the selected plateau
            \EndIf
            \State Obtain \(\widehat\alpha_{0.5}\), \(\widehat\xi\), and their fitted covariance \(\widehat V\); report the approximate 95\% Wald CI \(\widehat\xi\pm1.96\sqrt{\widehat V_{22}}\)
            \For{each requested design life \(T\)}
            \State Set \(b_T=\lceil\lambda T\rceil\) and compute \(\widehat D_{0.5}(T)\) and its pointwise 95\% CI using \eqref{eq:design_life_level_est} and \eqref{eq:design_life_ci}
            \EndFor
            \For{each higher quantile \(\tau'\in\{0.90,0.95,0.99\}\) requested for the applications}
            \State Compute \(\widehat Q_{\tau'}(M_b)\) on the selected plateau and estimate \(\widehat\alpha_{\tau'}\) using \eqref{eq:shared_xi_intercept}, holding \(\widehat\xi\) fixed
            \State Report \(\widehat D_{\tau'}(T)\) as point estimates on the same design-life grid
            \EndFor
        \end{algorithmic}
    
\end{algorithm}

\begin{algorithm}[htbp]
    \caption{Persistence workflow for pooled block-maxima EI}
    \label{alg:persistence-workflow}
    
        \begin{algorithmic}[1]
            \Require Series \(X_1,\ldots,X_N\), block-size grid \(\{b_k\}_{k=1}^{K}\), base estimator (Northrop or BB), block scheme (sliding or disjoint), and fitting method (OLS or FGLS)
            \State Compute the empirical distribution of the input series
            \For{each block size \(b_k\)}
            \State Compute \(\widehat\theta_{b_k}\) using the specified base estimator and block scheme
            \State Set \(z_{b_k}=\log(1/\widehat\theta_{b_k})\)
            \EndFor
            \State Select an EI stable window on the finite \(z_b\) path using the variance--roughness--curvature criterion and window-size weighting (\Cref{tab:implementation-defaults})
            \If{FGLS is selected}
            \State Estimate the transformed EI-path covariance by adaptive circular block bootstrap of the input series, recomputing the empirical distribution and EI path on each resample
            \State Regularize the selected covariance and fit the intercept-only model by pooled FGLS using \eqref{eq:pooled_z_gls}
            \Else
            \State Fit the intercept-only model by pooled OLS on the selected window
            \EndIf
            \State Obtain the unconstrained estimate \(\widetilde\mu\) and fitted standard error \(\widehat s_\mu\)
            \State Set \(\widehat\mu=\max\{0,\widetilde\mu\}\) and \(\widehat\theta=\exp(-\widehat\mu)\)
            \State Report \(\widehat\theta\), its approximate 95\% CI from \eqref{eq:theta_ci}, and the estimated limiting mean cluster size \(1/\widehat\theta\)
        \end{algorithmic}
    
\end{algorithm}

\section{Benchmark conventions and sensitivity analyses}

\subsection{Consolidated benchmark summary tables}\label{app:benchmark-core-tables}
\Cref{tab:benchmark-evi-summary,tab:benchmark-ei-summary} report consolidated benchmark summaries for the EVI and EI benchmarks.

\begin{table}[p]
\centering
\tiny
\setlength{\tabcolsep}{1pt}
\renewcommand{\arraystretch}{1.06}
\caption{EVI benchmark results for synthetic records of \(N=365\) observations from the Fréchet max-AR, moving-maxima \(q=99\), and Pareto additive AR(1) families. The parameter grid is \(\theta \in \{0.01, 0.10, 0.50, 1.0\}\) and \(\xi \in \{0.01, 0.03, 0.10, 0.30, 1.0, 3.0, 10.0\}\). Rows report methods and columns group results by process family and \(\theta\). Within each cell, the first line is the grid-average Winkler score (mean of scenario-level means) over the \(\xi\) grid; the second is the median of scenario-level median absolute percentage errors (APE, expressed as a fraction) over the same grid. All interval metrics use 95\% confidence intervals (\(\alpha = 0.05\)). Within each column, the unrounded minimum of each metric is bold, including exact ties.}
\label{tab:benchmark-evi-summary}
\begin{tabularx}{\textwidth}{>{\raggedright\arraybackslash}m{0.18\textwidth}>{\centering\arraybackslash}X>{\centering\arraybackslash}X>{\centering\arraybackslash}X>{\centering\arraybackslash}X>{\centering\arraybackslash}X>{\centering\arraybackslash}X>{\centering\arraybackslash}X>{\centering\arraybackslash}X>{\centering\arraybackslash}X>{\centering\arraybackslash}X>{\centering\arraybackslash}X>{\centering\arraybackslash}X}
\hline
method & \multicolumn{4}{c}{Fréchet max-AR} & \multicolumn{4}{c}{Moving maxima (\(q=99\))} & \multicolumn{4}{c}{Pareto additive AR(1)} \\
true $\theta$ & 0.01 & 0.10 & 0.50 & 1.0 & 0.01 & 0.10 & 0.50 & 1.0 & 0.01 & 0.10 & 0.50 & 1.0 \\
\hline
\shortstack[l]{median- \\ sliding-FGLS} & \shortstack[c]{\textbf{15.254} \\ 0.92} & \shortstack[c]{\textbf{8.317} \\ 0.55} & \shortstack[c]{5.382 \\ 0.39} & \shortstack[c]{5.444 \\ 0.37} & \shortstack[c]{\textbf{16.358} \\ 1.00} & \shortstack[c]{8.475 \\ 0.52} & \shortstack[c]{5.159 \\ 0.354} & \shortstack[c]{5.342 \\ 0.356} & \shortstack[c]{\textbf{19.732} \\ \textbf{0.93577}} & \shortstack[c]{\textbf{11.179} \\ 0.94} & \shortstack[c]{7.495 \\ 0.80} & \shortstack[c]{5.686 \\ 0.39} \\
\shortstack[l]{median- \\ sliding-OLS} & \shortstack[c]{74.191 \\ 0.92} & \shortstack[c]{39.550 \\ 0.55} & \shortstack[c]{27.149 \\ 0.38} & \shortstack[c]{23.983 \\ 0.36} & \shortstack[c]{82.514 \\ 1.00} & \shortstack[c]{38.958 \\ 0.52} & \shortstack[c]{23.730 \\ 0.375} & \shortstack[c]{23.130 \\ 0.348} & \shortstack[c]{75.583 \\ 0.93578} & \shortstack[c]{43.166 \\ 0.94} & \shortstack[c]{28.423 \\ 0.80} & \shortstack[c]{27.726 \\ 0.39} \\
\shortstack[l]{median- \\ disjoint-FGLS} & \shortstack[c]{16.157 \\ 0.93} & \shortstack[c]{8.598 \\ 0.54} & \shortstack[c]{5.627 \\ 0.41} & \shortstack[c]{5.712 \\ 0.36} & \shortstack[c]{16.636 \\ 1.00} & \shortstack[c]{\textbf{8.463} \\ 0.55} & \shortstack[c]{5.361 \\ 0.376} & \shortstack[c]{4.962 \\ 0.365} & \shortstack[c]{21.700 \\ 0.96540} & \shortstack[c]{11.797 \\ 0.96} & \shortstack[c]{9.254 \\ 0.80} & \shortstack[c]{6.223 \\ 0.38} \\
\shortstack[l]{median- \\ disjoint-OLS} & \shortstack[c]{69.869 \\ 0.92} & \shortstack[c]{31.504 \\ 0.55} & \shortstack[c]{21.210 \\ 0.39} & \shortstack[c]{23.596 \\ 0.35} & \shortstack[c]{77.852 \\ 1.00} & \shortstack[c]{34.752 \\ 0.55} & \shortstack[c]{22.133 \\ 0.373} & \shortstack[c]{18.397 \\ 0.340} & \shortstack[c]{70.590 \\ 0.96574} & \shortstack[c]{38.001 \\ 0.96} & \shortstack[c]{25.159 \\ 0.78} & \shortstack[c]{20.547 \\ 0.36} \\
\addlinespace[2pt]
\shortstack[l]{mean- \\ sliding-FGLS} & \shortstack[c]{48.947 \\ 0.90} & \shortstack[c]{25.978 \\ 0.42} & \shortstack[c]{22.033 \\ \textbf{0.14}} & \shortstack[c]{24.035 \\ 0.13} & \shortstack[c]{49.356 \\ 0.92} & \shortstack[c]{26.251 \\ 0.42} & \shortstack[c]{25.616 \\ 0.142} & \shortstack[c]{24.018 \\ 0.129} & \shortstack[c]{46.967 \\ 0.94099} & \shortstack[c]{28.670 \\ \textbf{0.92}} & \shortstack[c]{25.187 \\ 0.90} & \shortstack[c]{22.852 \\ 0.15} \\
\shortstack[l]{mean- \\ disjoint-OLS} & \shortstack[c]{74.644 \\ 0.92} & \shortstack[c]{64.553 \\ 0.50} & \shortstack[c]{63.979 \\ 0.19} & \shortstack[c]{63.623 \\ 0.17} & \shortstack[c]{78.342 \\ 0.94} & \shortstack[c]{65.258 \\ 0.49} & \shortstack[c]{63.433 \\ 0.188} & \shortstack[c]{63.678 \\ 0.173} & \shortstack[c]{76.223 \\ 0.96545} & \shortstack[c]{68.010 \\ 0.95} & \shortstack[c]{65.634 \\ 0.90} & \shortstack[c]{63.719 \\ 0.16} \\
\addlinespace[2pt]
\shortstack[l]{mode- \\ sliding-FGLS} & \shortstack[c]{39.336 \\ 0.91} & \shortstack[c]{9.605 \\ 0.68} & \shortstack[c]{10.569 \\ 0.63} & \shortstack[c]{10.338 \\ 0.67} & \shortstack[c]{24.247 \\ 1.00} & \shortstack[c]{8.630 \\ 0.66} & \shortstack[c]{12.141 \\ 0.619} & \shortstack[c]{12.567 \\ 0.640} & \shortstack[c]{37.405 \\ 0.96638} & \shortstack[c]{13.689 \\ 0.97} & \shortstack[c]{16.547 \\ 0.88} & \shortstack[c]{14.166 \\ 0.66} \\
\shortstack[l]{mode- \\ disjoint-OLS} & \shortstack[c]{66.095 \\ 0.90} & \shortstack[c]{35.617 \\ 0.58} & \shortstack[c]{25.017 \\ 0.46} & \shortstack[c]{28.516 \\ 0.44} & \shortstack[c]{78.657 \\ 1.00} & \shortstack[c]{29.672 \\ 0.54} & \shortstack[c]{24.555 \\ 0.435} & \shortstack[c]{29.213 \\ 0.423} & \shortstack[c]{68.723 \\ 0.99638} & \shortstack[c]{40.189 \\ 0.96} & \shortstack[c]{30.093 \\ 0.77} & \shortstack[c]{27.058 \\ 0.43} \\
\midrule
\shortstack[l]{Hill} & \shortstack[c]{75.279 \\ 0.94} & \shortstack[c]{18.888 \\ 0.42} & \shortstack[c]{3.773 \\ 0.18} & \shortstack[c]{\textbf{1.699} \\ \textbf{0.11}} & \shortstack[c]{82.103 \\ 1.00} & \shortstack[c]{18.935 \\ 0.42} & \shortstack[c]{3.541 \\ 0.161} & \shortstack[c]{\textbf{2.051} \\ \textbf{0.117}} & \shortstack[c]{75.554 \\ 0.94553} & \shortstack[c]{20.773 \\ 0.93} & \shortstack[c]{\textbf{5.338} \\ \textbf{0.66}} & \shortstack[c]{\textbf{1.734} \\ \textbf{0.11}} \\
\addlinespace[1.5pt]
\shortstack[l]{Max- \\ spectrum} & \shortstack[c]{63.854 \\ \textbf{0.85}} & \shortstack[c]{21.078 \\ \textbf{0.40}} & \shortstack[c]{6.964 \\ 0.15} & \shortstack[c]{4.896 \\ 0.12} & \shortstack[c]{63.122 \\ \textbf{0.86}} & \shortstack[c]{22.026 \\ \textbf{0.40}} & \shortstack[c]{6.653 \\ \textbf{0.139}} & \shortstack[c]{6.792 \\ 0.125} & \shortstack[c]{65.844 \\ 0.94097} & \shortstack[c]{25.506 \\ 0.94} & \shortstack[c]{8.961 \\ 0.69} & \shortstack[c]{6.604 \\ 0.13} \\
\addlinespace[1.5pt]
\shortstack[l]{Pickands} & \shortstack[c]{57.346 \\ 4.12} & \shortstack[c]{9.463 \\ 1.35} & \shortstack[c]{\textbf{3.003} \\ 0.62} & \shortstack[c]{3.023 \\ 0.78} & \shortstack[c]{93.836 \\ 5.47} & \shortstack[c]{10.598 \\ 1.36} & \shortstack[c]{\textbf{3.009} \\ 0.727} & \shortstack[c]{3.104 \\ 0.843} & \shortstack[c]{65.860 \\ 2.67430} & \shortstack[c]{18.354 \\ 2.59} & \shortstack[c]{6.705 \\ 1.20} & \shortstack[c]{2.506 \\ 0.50} \\
\addlinespace[1.5pt]
DEdH & \shortstack[c]{97.579 \\ 4.09} & \shortstack[c]{33.775 \\ 2.10} & \shortstack[c]{5.536 \\ 0.48} & \shortstack[c]{2.261 \\ 0.30} & \shortstack[c]{150.732 \\ 2.15} & \shortstack[c]{34.711 \\ 1.98} & \shortstack[c]{6.052 \\ 0.477} & \shortstack[c]{2.961 \\ 0.299} & \shortstack[c]{94.024 \\ 2.35465} & \shortstack[c]{34.865 \\ 1.78} & \shortstack[c]{7.157 \\ 0.97} & \shortstack[c]{2.638 \\ 0.31} \\
\hline
\end{tabularx}
\end{table}

\begin{table}[p]
\centering
\tiny
\setlength{\tabcolsep}{1pt}
\renewcommand{\arraystretch}{1.06}
\caption{EI benchmark results for synthetic records of \(N=365\) observations from the Fréchet max-AR, moving-maxima \(q=99\), and Pareto additive AR(1) families. The parameter grid is \(\xi \in \{0.01, 0.50, 1.0, 5.0\}\) and \(\theta \in \{0.10, 0.15, 0.25, 0.40, 0.60, 0.80, 1.0\}\). Rows report methods and columns group results by process family and \(\xi\). Within each cell, the first line is the grid-average Winkler score (mean of scenario-level means) over the \(\theta\) grid; the second is the median of scenario-level median absolute percentage errors (APE, expressed as a fraction) over the same grid. All interval metrics use 95\% confidence intervals (\(\alpha = 0.05\)). Within each column, the unrounded minimum of each metric is bold, including exact ties. An FGLS attempt with degenerate bootstrap covariance is retained as noncoverage; interval-score and point-error summaries are conditional on successful fits.}
\label{tab:benchmark-ei-summary}
\begin{tabularx}{\textwidth}{>{\raggedright\arraybackslash}m{0.18\textwidth}>{\centering\arraybackslash}X>{\centering\arraybackslash}X>{\centering\arraybackslash}X>{\centering\arraybackslash}X>{\centering\arraybackslash}X>{\centering\arraybackslash}X>{\centering\arraybackslash}X>{\centering\arraybackslash}X>{\centering\arraybackslash}X>{\centering\arraybackslash}X>{\centering\arraybackslash}X>{\centering\arraybackslash}X}
\hline
method & \multicolumn{4}{c}{Fréchet max-AR} & \multicolumn{4}{c}{Moving maxima (\(q=99\))} & \multicolumn{4}{c}{Pareto additive AR(1)} \\
true $\xi$ & 0.01 & 0.50 & 1.0 & 5.0 & 0.01 & 0.50 & 1.0 & 5.0 & 0.01 & 0.50 & 1.0 & 5.0 \\
\hline
\shortstack[l]{Northrop- \\ sliding-FGLS} & \shortstack[c]{1.023 \\ 0.18} & \shortstack[c]{1.060 \\ 0.19} & \shortstack[c]{0.953 \\ 0.16} & \shortstack[c]{1.027 \\ 0.121} & \shortstack[c]{0.947 \\ 0.154} & \shortstack[c]{0.925 \\ 0.17} & \shortstack[c]{0.903 \\ 0.159} & \shortstack[c]{1.056 \\ 0.17} & \shortstack[c]{9.072 \\ 0.66} & \shortstack[c]{1.242 \\ 0.2233} & \shortstack[c]{0.584 \\ 0.1071} & \shortstack[c]{0.984 \\ 0.16} \\
\shortstack[l]{Northrop- \\ sliding-OLS} & \shortstack[c]{2.585 \\ 0.18} & \shortstack[c]{2.624 \\ 0.17} & \shortstack[c]{2.422 \\ 0.15} & \shortstack[c]{2.447 \\ \textbf{0.117}} & \shortstack[c]{2.456 \\ 0.150} & \shortstack[c]{2.410 \\ 0.17} & \shortstack[c]{2.341 \\ 0.153} & \shortstack[c]{2.516 \\ 0.16} & \shortstack[c]{9.999 \\ 0.70} & \shortstack[c]{3.010 \\ 0.2235} & \shortstack[c]{1.971 \\ 0.1053} & \shortstack[c]{2.411 \\ 0.16} \\
\shortstack[l]{Northrop- \\ disjoint-FGLS} & \shortstack[c]{1.332 \\ 0.21} & \shortstack[c]{1.399 \\ 0.18} & \shortstack[c]{1.336 \\ 0.17} & \shortstack[c]{1.388 \\ 0.158} & \shortstack[c]{1.321 \\ 0.157} & \shortstack[c]{1.319 \\ 0.20} & \shortstack[c]{1.227 \\ 0.144} & \shortstack[c]{1.333 \\ 0.20} & \shortstack[c]{8.988 \\ 0.67} & \shortstack[c]{1.342 \\ 0.2303} & \shortstack[c]{0.899 \\ 0.1094} & \shortstack[c]{1.296 \\ 0.18} \\
\shortstack[l]{Northrop- \\ disjoint-OLS} & \shortstack[c]{2.394 \\ 0.21} & \shortstack[c]{2.494 \\ 0.19} & \shortstack[c]{2.315 \\ 0.16} & \shortstack[c]{2.398 \\ 0.153} & \shortstack[c]{2.309 \\ 0.156} & \shortstack[c]{2.305 \\ 0.20} & \shortstack[c]{2.200 \\ 0.140} & \shortstack[c]{2.381 \\ 0.18} & \shortstack[c]{9.505 \\ 0.69} & \shortstack[c]{2.566 \\ 0.2231} & \shortstack[c]{1.849 \\ 0.1115} & \shortstack[c]{2.338 \\ 0.17} \\
\addlinespace[2pt]
\shortstack[l]{BB- \\ sliding-FGLS} & \shortstack[c]{0.982 \\ 0.19} & \shortstack[c]{1.040 \\ 0.18} & \shortstack[c]{0.906 \\ 0.15} & \shortstack[c]{0.978 \\ 0.131} & \shortstack[c]{0.911 \\ 0.150} & \shortstack[c]{0.889 \\ 0.16} & \shortstack[c]{0.878 \\ 0.150} & \shortstack[c]{1.017 \\ 0.17} & \shortstack[c]{9.075 \\ 0.67} & \shortstack[c]{\textbf{1.119} \\ \textbf{0.1927}} & \shortstack[c]{\textbf{0.561} \\ \textbf{0.0865}} & \shortstack[c]{0.930 \\ 0.17} \\
\shortstack[l]{BB- \\ sliding-OLS} & \shortstack[c]{2.550 \\ 0.18} & \shortstack[c]{2.608 \\ 0.17} & \shortstack[c]{2.383 \\ 0.15} & \shortstack[c]{2.421 \\ 0.127} & \shortstack[c]{2.431 \\ \textbf{0.147}} & \shortstack[c]{2.388 \\ 0.16} & \shortstack[c]{2.329 \\ \textbf{0.138}} & \shortstack[c]{2.518 \\ 0.15} & \shortstack[c]{10.015 \\ 0.70} & \shortstack[c]{2.880 \\ 0.1973} & \shortstack[c]{1.947 \\ 0.0924} & \shortstack[c]{2.358 \\ 0.17} \\
\shortstack[l]{BB- \\ disjoint-FGLS} & \shortstack[c]{1.322 \\ 0.22} & \shortstack[c]{1.391 \\ 0.19} & \shortstack[c]{1.352 \\ 0.18} & \shortstack[c]{1.336 \\ 0.168} & \shortstack[c]{1.308 \\ 0.157} & \shortstack[c]{1.297 \\ 0.19} & \shortstack[c]{1.206 \\ 0.150} & \shortstack[c]{1.376 \\ 0.19} & \shortstack[c]{8.939 \\ 0.67} & \shortstack[c]{1.308 \\ 0.2218} & \shortstack[c]{0.916 \\ 0.1016} & \shortstack[c]{1.273 \\ 0.18} \\
\shortstack[l]{BB- \\ disjoint-OLS} & \shortstack[c]{2.388 \\ 0.21} & \shortstack[c]{2.480 \\ 0.18} & \shortstack[c]{2.319 \\ 0.15} & \shortstack[c]{2.366 \\ 0.154} & \shortstack[c]{2.306 \\ 0.150} & \shortstack[c]{2.304 \\ 0.18} & \shortstack[c]{2.200 \\ 0.141} & \shortstack[c]{2.400 \\ 0.18} & \shortstack[c]{9.548 \\ 0.68} & \shortstack[c]{2.539 \\ 0.2188} & \shortstack[c]{1.853 \\ 0.0997} & \shortstack[c]{2.285 \\ 0.17} \\
\midrule
\shortstack[l]{Ferro- \\ Segers} & \shortstack[c]{1.768 \\ 0.18} & \shortstack[c]{1.698 \\ 0.19} & \shortstack[c]{1.674 \\ 0.20} & \shortstack[c]{1.684 \\ 0.227} & \shortstack[c]{1.615 \\ 0.242} & \shortstack[c]{1.813 \\ 0.18} & \shortstack[c]{1.400 \\ 0.197} & \shortstack[c]{1.302 \\ 0.22} & \shortstack[c]{16.406 \\ 0.79} & \shortstack[c]{1.686 \\ 0.2054} & \shortstack[c]{1.768 \\ 0.1620} & \shortstack[c]{1.441 \\ 0.22} \\
\addlinespace[1.5pt]
\shortstack[l]{K-gaps} & \shortstack[c]{\textbf{0.492} \\ \textbf{0.12}} & \shortstack[c]{\textbf{0.512} \\ \textbf{0.12}} & \shortstack[c]{\textbf{0.547} \\ \textbf{0.12}} & \shortstack[c]{\textbf{0.476} \\ 0.123} & \shortstack[c]{\textbf{0.518} \\ 0.162} & \shortstack[c]{\textbf{0.539} \\ \textbf{0.14}} & \shortstack[c]{\textbf{0.458} \\ 0.144} & \shortstack[c]{\textbf{0.553} \\ \textbf{0.13}} & \shortstack[c]{\textbf{5.554} \\ \textbf{0.63}} & \shortstack[c]{1.180 \\ 0.2171} & \shortstack[c]{0.635 \\ 0.1269} & \shortstack[c]{\textbf{0.502} \\ \textbf{0.11}} \\
\addlinespace[1.5pt]
\shortstack[l]{Northrop\\(fixed-\(b\))} & \shortstack[c]{1.570 \\ 0.18} & \shortstack[c]{1.614 \\ 0.19} & \shortstack[c]{1.472 \\ 0.16} & \shortstack[c]{1.502 \\ 0.125} & \shortstack[c]{1.481 \\ 0.159} & \shortstack[c]{1.444 \\ 0.18} & \shortstack[c]{1.419 \\ 0.161} & \shortstack[c]{1.573 \\ 0.18} & \shortstack[c]{9.409 \\ 0.66} & \shortstack[c]{1.985 \\ 0.2224} & \shortstack[c]{1.032 \\ 0.1064} & \shortstack[c]{1.515 \\ 0.17} \\
\addlinespace[1.5pt]
\shortstack[l]{BB\\(fixed-\(b\))} & \shortstack[c]{1.473 \\ 0.19} & \shortstack[c]{1.525 \\ 0.19} & \shortstack[c]{1.363 \\ 0.16} & \shortstack[c]{1.414 \\ 0.131} & \shortstack[c]{1.396 \\ 0.161} & \shortstack[c]{1.352 \\ 0.17} & \shortstack[c]{1.333 \\ 0.156} & \shortstack[c]{1.488 \\ 0.18} & \shortstack[c]{9.369 \\ 0.67} & \shortstack[c]{1.818 \\ 0.1934} & \shortstack[c]{0.992 \\ 0.0874} & \shortstack[c]{1.406 \\ 0.18} \\
\hline
\end{tabularx}
\end{table}

\subsection{External baselines: definitions, tuning rules, and intervals}\label{app:benchmark-baselines}
External baselines are evaluated over pre-specified grids of candidate thresholds or block sizes.

For the raw-sample EVI estimators (Hill, Pickands, and DEdH), positive finite observations are sorted in descending order.
Each estimator is evaluated over a log-spaced tail-count grid \(k\in[8,\lfloor 0.25n\rfloor]\) with 24 candidate values.
Here \(n\) is the number of retained positive finite observations and \(k\) is the tail-count tuning parameter.
After removing nonfinite estimates, we score each window of four consecutive candidates along the \(k\)-path by the mean squared deviation of its estimates from their mean plus half the mean absolute second difference.
    We select the central candidate of the lowest-scoring window, taking the lower of the two central candidates for even window lengths; score ties favor earlier windows.
    A path shorter than the window length is used in full.
Intervals are reported using each estimator's native asymptotic Gaussian or Wald construction \citep{hill_simple_1975,pickands_1975,dekkers_einmahl_dehaan_1989,fedotenkov_review_2020}.

The max-spectrum baseline is evaluated on its native dyadic scale.
Block sizes are \(2^j\), restricted to scales with at least two complete blocks, using an entirely finite input series in its original time order; the resulting block maxima must be positive before the log transformation.
The start-scale path uses the variance--curvature score and central-candidate selection rule above, with three-point windows.
    We report the native weighted slope estimator of \citet{stoev_michailidis_2010} with a heteroskedasticity-robust Wald interval.

For the threshold-based EI baselines, the candidate quantile levels are fixed at \(u\in\{0.90,0.95\}\).
Here \(u\) labels a marginal quantile level; the threshold is the corresponding empirical quantile.
Ferro--Segers is fitted at both thresholds.
    When both estimates are available, we select \(u=0.90\) if their 95\% confidence intervals overlap and \(u=0.95\) otherwise; a sole available estimate is retained.
K-gaps uses the same threshold set with run parameters \(K\in\{1,2\}\) in observation steps.
    At each threshold, overlapping 95\% confidence intervals favor \(K=1\); otherwise, \(K=2\) is selected.
    The threshold is then chosen by the same interval-overlap rule.

The external EI comparison also includes native fixed-\(b\) block-maxima baselines.
For reference, the Northrop log pseudo-likelihood can be written schematically as
\begin{equation}\label{eq:northrop_profile}
    \ell_{\mathrm N}(\theta)
    =
    m\log \theta - \theta \sum_{j=1}^{m} T_j,
\end{equation}
for positive path statistics \(T_1,\ldots,T_m\) at the selected block size.
Here \(m\) is the number of extracted block maxima \(V_1,\ldots,V_m\), and \(T_j=-b\log\widehat F(V_j)\).
    The implementation uses the scaled-rank marginal estimate \(\widehat F(x)=(N+1)^{-1}\sum_{t=1}^N\mathbf{1}\{X_t\le x\}\) from the full retained series, where \(\mathbf{1}\) denotes the indicator function.
For the BB fixed-\(b\) approximation, the point estimate takes the form
\begin{equation}\label{eq:bb_fixed_b}
    \widehat\theta_{\mathrm{BB}}
    =
    \max\left\{\frac{1}{\bar S_b} - \frac{1}{b},\,0\right\},
\end{equation}
where \(\bar S_b\) is the mean rolling-minimum statistic at the chosen block size.
With the same block maxima and marginal estimate, \(\bar S_b=m^{-1}\sum_{j=1}^m b\{1-\widehat F(V_j)\}\).
    The implementation caps EI estimates at one and uses a positive numerical floor before logarithmic transformation.
The fixed-\(b\) Northrop and BB estimators provide the estimates at individual block sizes that form the persistence paths.
    The proposed workflow selects a stable window on each transformed path and pools its retained values using OLS or covariance-aware FGLS.

The external EI baselines retain their native intervals: bounded Wald intervals for Ferro--Segers \citep{ferro_segers_2003}, profile-likelihood intervals for K-gaps \citep{suveges_davison_2010}, adjusted profile-likelihood intervals for Northrop's estimator \citep{northrop_efficient_2015,chandler_bate_2007}, and
bounded Wald intervals for the BB estimator \citep{berghaus_weak_2018}.
In the implementation, profile-likelihood endpoints are obtained by bracketed scalar root search \citep{oliveira_enhancement_2020}.

\subsection{Shrinkage sensitivity for covariance-regularized FGLS}\label{app:shrinkage-sensitivity}
\begin{figure}[htbp]
    \centering
    \includegraphics[width=0.98\textwidth]{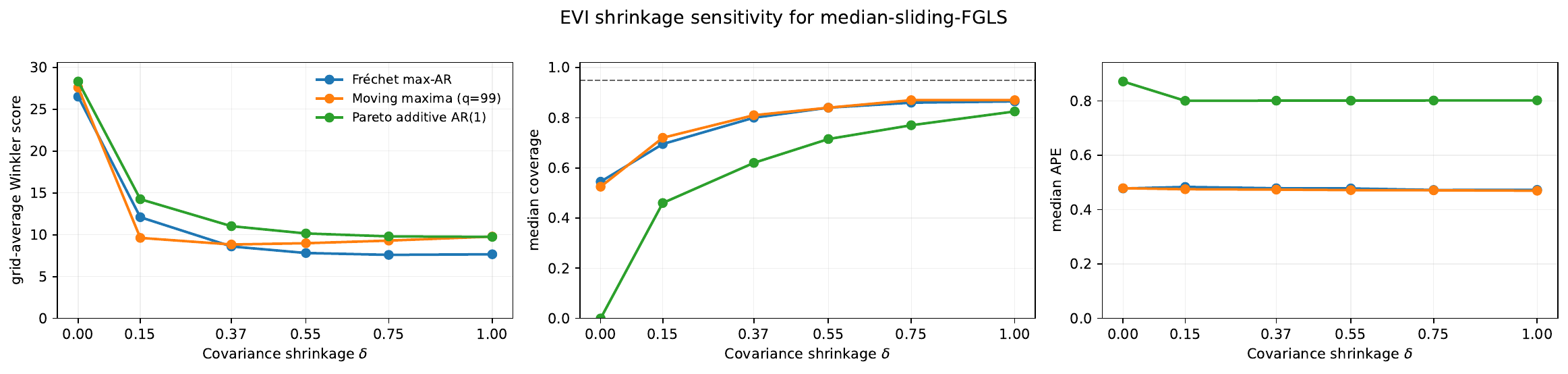}
    \caption{
        EVI shrinkage sensitivity for the median-sliding-FGLS workflow.
        Panels report grid-average Winkler score, median empirical coverage, and median APE over the EVI parameter grid within each process family as the covariance-shrinkage factor \(\delta\) varies over a fixed grid.
        The benchmark default is \(\delta=0.37\).
    }
    \label{fig:benchmark-evi-shrinkage}
\end{figure}

\begin{figure}[htbp]
    \centering
    \includegraphics[width=0.98\textwidth]{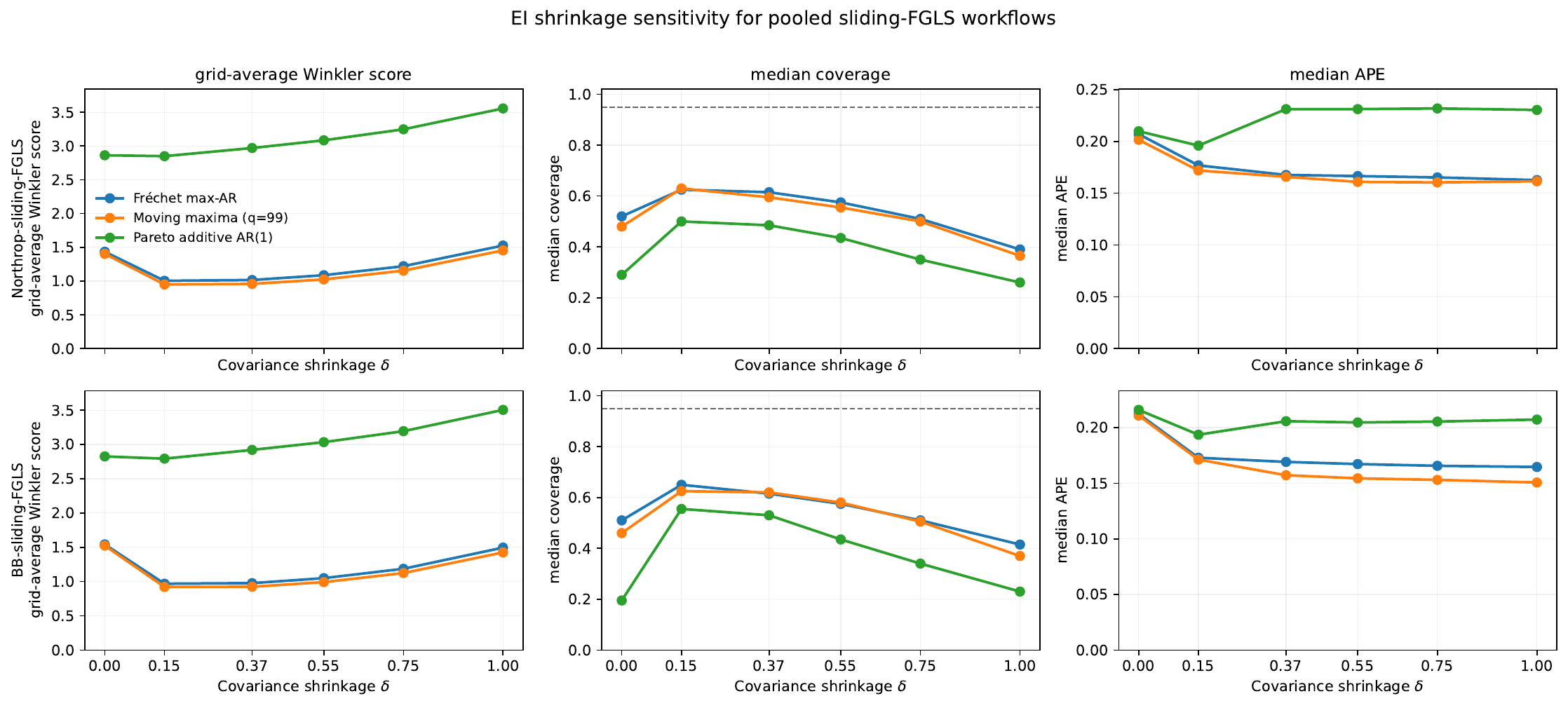}
    \caption{
        EI shrinkage sensitivity for the pooled sliding-FGLS workflows.
        Rows correspond to Northrop-sliding-FGLS and BB-sliding-FGLS.
        Panels report grid-average Winkler score, median empirical coverage, and median APE over the EI parameter grid within each process family as \(\delta\) varies over a fixed grid.
        The benchmark default is \(\delta=0.37\); failed covariance fits count as noncoverage and have no finite score.
    }
    \label{fig:benchmark-ei-shrinkage}
\end{figure}

The severity and persistence benchmarks use a common empirical diagonal shrinkage factor \(\delta=0.37\) to regularize bootstrap covariance estimates before FGLS inversion.
The sensitivity summaries in \Cref{fig:benchmark-evi-shrinkage,fig:benchmark-ei-shrinkage} vary only \(\delta\) over the fixed grid \(\{0.00,0.15,0.37,0.55,0.75,1.00\}\), using the same 84 scenarios, record length \(N=365\), and 100 simulation replicates per scenario as each main branch.
    Each fit retains its original-data window and the bootstrap replicates obtained by adaptive stopping at the default \(\delta=0.37\); these replicates are reused for all other shrinkage values.

For the severity branch, moving from no shrinkage to \(\delta=0.37\) reduces the grid-average Winkler score from \(27.47\) to \(9.49\).
    Larger shrinkage can improve the grid-average Winkler score further: \(\delta=0.75\) gives \(8.90\), but the process-family minima differ, occurring at \(0.75\) for Fr\'echet max-AR, \(0.37\) for moving maxima, and \(1.00\) for Pareto additive AR(1) on this grid.
    Median APE changes little among the positive shrinkage values, while median coverage generally increases with shrinkage without reaching the nominal 95\% level.
For the persistence branch, both sliding-FGLS methods have their lowest grid-average Winkler scores at \(\delta=0.15\): \(1.561\) for BB and \(1.601\) for Northrop, compared with \(1.607\) and \(1.648\) at the fixed default.
    Full diagonal shrinkage worsens these scores to \(2.141\) and \(2.179\), respectively
.

\section{Application data and supplementary analyses}\label{app:application-uncertainty}

\subsection{Data preparation and bootstrap replication}\label{app:application-computation}
The retained streamflow records contain 37,135 daily mean discharge observations at USGS 08066500 (1 May 1924--31 December 2025) and 15,189 at USGS 02366500 (1 June 1984--31 December 2025).
    Although the Florida archive begins on 1 October 1930, its longest gap-free record ending at the cutoff begins on 1 June 1984.
    Both EVI and EI use these retained records.

Each NFIP calendar series contains 17,532 days from 1 January 1978 to 31 December 2025.
    The active-day building-claim series contains 5,807 observations in Texas (9 January 1978--14 December 2025) and 5,440 in Florida (8 January 1978--11 December 2025).
    These endpoints identify the first and last positive loss dates; active-day rates use the full 48-year analysis period.

Inflation adjustment uses the monthly CPI-U series CUUR0000SA0 \citep{bls_cpi_u_historical}, with the official 2025 annual average, \(321.943\), as the constant-dollar base \citep{bls_cpi_u_annual}.
    The unavailable October 2025 value is imputed as the geometric mean of September and November, \(324.461\), and flagged in the CPI-U input.

The adaptive stopping rule in \ref{app:implementation-defaults} yields \(R=256\) for both streamflow severity fits, \(128\) for Texas NFIP, and \(512\) for Florida NFIP.
    All four BB-sliding-FGLS EI fits use \(R=256\); Northrop-sliding-FGLS uses \(R=512\) for Florida streamflow and \(256\) for the other cases.

\subsection{Window sensitivity and extrapolation}
We assess window-selection sensitivity by refitting the EVI estimator and BB-sliding-FGLS over their three best-ranked admissible windows under the fixed selection rules (\Cref{tab:application-selection-sensitivity}).

After trimming 15\% of the path points from each end, we rank all contiguous windows containing at least five points for EVI or four for EI.
    Plateaus use \eqref{eq:plateau_score}; EI stable windows use the variance--roughness--curvature criterion on the transformed \(z_b\) path, with settings given in \Cref{tab:implementation-defaults}.
    Both scores divide their numerator by the square root of the number of points in the candidate window.
    The reported min--max ranges summarize sensitivity to window choice and complement the conditional intervals in \Cref{tab:application-summary}; they are not post-selection confidence intervals.

The four BB-sliding-FGLS EI estimates vary little across the three best-ranked windows.
    EVI sensitivity is greatest for Texas NFIP claims, while the NFIP EVI estimates remain above those for streamflow.

\Cref{tab:application-extrapolation} reports the selected plateaus, the block sizes for the 10- and 50-year design-life levels, and their ratios to the largest fitted block size.

\begin{table}[htbp]
\centering
\caption{Window-selection sensitivity of EVI and EI estimates. Each cell reports the selected estimate and the min--max range over the three best-ranked admissible plateaus (EVI) or EI stable windows under the fixed selection rules. EVI uses median-sliding-FGLS and EI uses BB-sliding-FGLS. These ranges describe sensitivity to window choice and complement the conditional intervals in \Cref{tab:application-summary}.}
\label{tab:application-selection-sensitivity}
\begin{tabular}{llc}
\hline
Application & $\xi$ [range] & $\theta$ [range] \\
\hline
Texas streamflow & 0.64 [0.59, 0.64] & 0.0488 [0.0488, 0.0495] \\
Florida streamflow & 0.33 [0.32, 0.33] & 0.0551 [0.0551, 0.0569] \\
Texas NFIP claims & 1.40 [1.31, 1.49] & 0.3118 [0.3117, 0.3129] \\
Florida NFIP claims & 1.38 [1.37, 1.43] & 0.3092 [0.3087, 0.3092] \\
\hline
\end{tabular}
\end{table}

\begin{table}[htbp]
\centering
\scriptsize
\setlength{\tabcolsep}{4pt}
\caption{Selected plateaus and design-life block sizes for the four application cases. The plateau column reports the block-size range retained for the median-sliding-FGLS fit, with maximum block size \(b_{\max,\mathrm{fit}}\). The \(b_{10}\) and \(b_{50}\) columns give the block sizes for 10- and 50-year design lives on the application-specific observation clock. The ratios compare each design-life block size with the largest fitted block size.}
\label{tab:application-extrapolation}
\begin{tabular}{p{0.23\textwidth}p{0.17\textwidth}rrrrr}
\toprule
Application & Clock & \shortstack[c]{Plateau} & \(b_{10}\) & \(b_{50}\) & \(\frac{b_{10}}{b_{\max,\mathrm{fit}}}\) & \(\frac{b_{50}}{b_{\max,\mathrm{fit}}}\) \\
\midrule
Texas streamflow & calendar day & 14--183 & 3,653 & 18,263 & 20.0 & 99.8 \\
Florida streamflow & calendar day & 17--26 & 3,653 & 18,263 & 140 & 702 \\
Texas NFIP claims & active day & 14--21 & 1,210 & 6,049 & 57.6 & 288 \\
Florida NFIP claims & active day & 10--19 & 1,134 & 5,667 & 59.7 & 298 \\
\bottomrule
\end{tabular}
\end{table}

\subsection{Streamflow GEV scale comparison}
We fit GEV distributions by L-moments to calendar-year (January--December) maxima of daily mean discharge \citep{hosking_lmoments_1990}.
    The same continuous records used by UniBM provide 101 annual maxima for Texas (1925--2025) and 41 for Florida (1985--2025).
    All retained years have complete daily coverage and meet the pre-specified 97\% requirement; the incomplete initial years, 1924 and 1984, are excluded.

\Cref{tab:application-streamflow-gev-check} compares GEV return levels, which are upper quantiles of annual maxima, with UniBM median design-life levels on the calendar-day clock.
    The table therefore compares different probability targets and serves as a hydrologic scale check.

The fitted GEV shapes are approximately \(-0.03\) for Texas and \(0.29\) for Florida, compared with UniBM slopes of \(0.64\) and \(0.33\), respectively.
    The Texas difference calls for assessment of tail-index stability across block sizes and estimation methods.
    In simulated precipitation, \citet{ben_alaya_block_maximum_2020} showed that block-length dependence of fitted GEV shapes can bias long-period return-level estimates.

\begin{table}[htbp]
\centering
\tiny
\caption{Streamflow scale comparison against conventional calendar-year-maxima GEV return levels. GEV entries are L-moment point estimates from January--December years meeting the predeclared 97 percent daily-coverage gate; retained years may have up to 3 percent missing daily observations \citep{hosking_lmoments_1990}; UniBM entries are median design-life levels with conditional 95 percent confidence intervals from the selected fit. The entries represent different quantiles and do not validate one another. This distinction alone does not explain the large Texas discrepancy, which cautions against extrapolating the selected finite-block scaling relation. GEV confidence intervals are not computed here, so the comparison is not a formal test of agreement. All discharge entries are in \(10^3\,\mathrm{ft}^3\,\mathrm{s}^{-1}\).}
\label{tab:application-streamflow-gev-check}
\setlength{\tabcolsep}{1pt}
\begin{tabular}{p{0.16\textwidth}p{0.10\textwidth}p{0.15\textwidth}p{0.20\textwidth}p{0.15\textwidth}p{0.20\textwidth}}
\hline
Application & \shortstack[c]{Calendar-year\\ maxima\\ count} & \shortstack[c]{GEV 10y\\ return level} & \shortstack[c]{UniBM 10y\\ design-life level} & \shortstack[c]{GEV 50y\\ return level} & \shortstack[c]{UniBM 50y\\ design-life level} \\
\hline
Texas streamflow & 101 & 84.2 & 232 [193, 280] & 116 & 655 [503, 854] \\
Florida streamflow & 41 & 71.6 & 38.5 [9.64, 154] & 133 & 65.0 [10.6, 398] \\
\hline
\end{tabular}
\end{table}

\subsection{USGS gauge screening}
\Cref{tab:application-usgs-screening} summarizes the candidate gauges and screening results for each state; the table caption specifies the eligibility criteria and ranking order.
    Record lengths refer to the gap-free daily records retained for screening, all ending on 31 December 2025.
    The selected gauges are Trinity River at Romayor (USGS 08066500) in Texas and Choctawhatchee River near Bruce (USGS 02366500) in Florida.
    Both meet the eligibility criteria and have positive lower confidence limits for the EVI in the screening fits.

\begin{table}[htbp]
\centering
\tiny
\caption{USGS streamflow candidate pools for the reported streamflow applications. Sites were screened with pre-specified criteria: minimum record length 20 years, minimum plateau size 5 block-size points, lower confidence limit for \(\xi\) at least -0.25, and positive-maxima share within that window at least 0.95. Ranking then prioritizes Fréchet-domain support, number of plateau points, record length, and lower confidence limit for \(\xi\).}
\label{tab:application-usgs-screening}
\setlength{\tabcolsep}{1pt}
\begin{tabular}{p{0.12\textwidth}p{0.23\textwidth}p{0.08\textwidth}p{0.15\textwidth}p{0.08\textwidth}p{0.10\textwidth}p{0.10\textwidth}}
\hline
\shortstack[c]{State /\\ site} & Station & Chosen & Fréchet support & Window points & Record years & \shortstack[c]{$\xi$ lower\\CI limit} \\
\hline
FL 02366500 & Choctawhatchee River near Bruce & yes & yes & 5 & 41.6 & 0.04 \\
FL 02236000 & St. Johns River near Deland & no & no & 7 & 0.7 & -0.50 \\
FL 02320500 & Suwannee River at Branford & no & no & 5 & 94.5 & -0.06 \\
TX 08066500 & Trinity River at Romayor & yes & yes & 23 & 101.7 & 0.60 \\
TX 08114000 & Brazos River at Richmond & no & yes & 5 & 103.2 & 0.48 \\
TX 08158000 & Colorado River at Austin & no & no & 6 & 8.9 & -0.47 \\
\hline
\end{tabular}
\end{table}

\end{document}